\documentclass[fleqn,11pt]{article}
\usepackage{amsfonts,amsbsy,amsmath,amssymb,
array,graphicx,latexsym,booktabs,enumerate,amsthm,enumerate}
\usepackage{colortbl}
\usepackage{natbib}

\allowdisplaybreaks

\newtheorem{thm}{Theorem}
\newtheorem{assumption}{Assumption}

\newtheorem{lemma}{Lemma}

\numberwithin{algo}{section}

\newtheorem{algorithm}{Algorithm}

\numberwithin{equation}{section}
\numberwithin{thm}{section}
\numberwithin{lemma}{section}
\numberwithin{assumption}{section}

\numberwithin{equation}{section}

\begin{document}

\setcounter{page}{1} 
\begin{center}
\textbf{\LARGE Testing Monotonicity of Mean Potential Outcomes in a Continuous Treatment with High-Dimensional Data}
\end{center}

\begin{center}
\vspace*{15mm}

\textbf{\large Yu-Chin Hsu$^{\dag}$}{\large{} }

\vspace*{2mm}
Institute of Economics, Academia Sinica\\
Department of Finance, National Central University\\
Department of Economics, National Chengchi University and\\
CRETA, National Taiwan University

\vspace*{3mm} \textbf{\large Martin Huber$^{*}$}

\vspace*{2mm} Department of Economics, University of Fribourg

\vspace*{3mm} \textbf{\large Ying-Ying Lee$^{\ddag}$}

\vspace*{2mm} Department of Economics, University of California, Irvine

\vspace*{3mm} \textbf{\large Chu-An Liu$^{\S}$}

\vspace*{2mm} Institute of Economics, Academia Sinica

\vspace*{5mm}
This version: August 4, 2022
\end{center}

\vfill{} \noindent  $^{\dag}$ {\footnotesize ychsu@econ.sinica.edu.tw}, $^{*}$ {\footnotesize martin.huber@unifr.ch}, $^{\ddag}$ {\footnotesize yingying.lee@uci.edu}, $^{\S}$ {\footnotesize caliu@econ.sinica.edu.tw.  }

\noindent {\footnotesize Acknowledgments: Yu-Chin Hsu gratefully acknowledges
research support from the National Science and Technology Council of Taiwan (NSTC111-2628-H-001-001), the Academia Sinica Investigator Award of the Academia Sinica, Taiwan (AS-IA-110-H01),
and the Center for Research in Econometric Theory and Applications (107L9002) from the
Featured Areas Research Center Program within the framework of the Higher Education
Sprout Project by the Ministry of Education of Taiwan. Chu-An Liu gratefully acknowledges research support from the Academia Sinica Career Development Award (AS-CDA-110-H02).}

\thispagestyle{empty}
\newpage

\vspace*{30mm}

\begin{center}
 {\Large \bf Abstract}
\end{center}

While most treatment evaluations focus on binary interventions, a growing literature also considers continuously distributed treatments. 
We propose a Cram\'{e}r-von Mises-type test for testing whether the mean potential outcome given a specific treatment has a weakly monotonic relationship with the treatment dose under a weak unconfoundedness assumption.
In a nonseparable structural model, applying our method amounts to testing monotonicity of the average structural function in the continuous treatment of interest.
To flexibly control for a possibly high-dimensional set of covariates in our testing approach, we propose a double debiased machine learning estimator that accounts for covariates in a data-driven way.
We show that the proposed test controls asymptotic size and is consistent against any fixed alternative. These theoretical findings are supported by the Monte-Carlo simulations. As an empirical illustration, we apply our test to the Job Corps study and reject a weakly negative relationship between the treatment (hours in academic and vocational training) and labor market performance among relatively low treatment values.

\vspace*{5mm}

\vfill \noindent


\thispagestyle{empty}
\pagebreak
\setcounter{page}{1}

\section{Introduction}
Even though many studies on treatment or policy evaluation investigate the effects of binary or discrete interventions, a growing literature also considers the assessment of continuously distributed treatments, e.g.\ hours spent in a training program whose effect on labor market performance is of interest.  Most contributions like \cite{Imbens2000}, \cite{HiranoImbens2004}, \cite{Flores2007}, \cite{Floresetal2012}, \cite{GalvaoWang2015}, \cite{Lee2018} and \cite{CL} focus on the identification and estimation of the average dose-response function (ADF), which corresponds to the mean potential outcome as a function of the treatment dose.  This permits assessing the average treatment effect (ATE) as the difference in the ADF assessed at two distinct treatment doses of interest, while \cite{HiranoImbens2004}, \cite{Floresetal2012}, and \cite{CL} also consider the marginal effect of slightly increasing the treatment dose, which is the derivative of the ADF. Rather than considering the total effect of the treatment, \cite{Huberetal2020} suggest a causal mediation approach to disentangle the ATE into its direct effect and indirect effect operating through intermediate variables or mediators to assess the causal mechanisms of the treatment.

In this paper, we propose a method for testing whether the ADF has a weakly monotonic relationship with (i.e.\ is weakly increasing or decreasing in) the treatment dose under a weak unconfoundedness assumption, implying that confounder of the treatment-outcome relation can be controlled for by observed covariates. Such a test appears interesting for verifying shape restrictions, e.g.\ whether increasing the treatment dose always has a non-negative effect, no matter what the baseline level of treatment is. Moreover, the treatment effect model is known to be equivalent to a nonseparable structural model of a nonseparable outcome with a general disturbance, as for instance \cite{IN09ETA} and \cite{Lee2018}. In this case, the ADF corresponds to the average structural function in \cite{BP03}.
Therefore our test can be applied to testing monotonicity of the average structural function in a nonseparable structural model under a conditional independence assumption.

To construct our test, we first transform the null hypothesis of a monotonic relationship to countably many moment inequalities based on the generalized instrumental function approach of \cite{HsuLiuShi2019} and \cite{HsuShen2020}.  We construct a Cram\'{e}r-von Mises-type test statistic based on the estimated moments, which are shown to converge to a Gaussian process at the parametric regular root-$n$ rate. Importantly, by making use of moment inequalities, our method does not rely on the nonparametric estimation of the ADF or the marginal effects, which would converge at slower nonparametric rates. To compute the critical value for our test, we apply a multiplier bootstrap method and the generalized moment selection (GMS) approach of \cite{AndrewsShi2013, AndrewsShi2014}.  We demonstrate that our test controls asymptotic size and is consistent against any fixed alternative.

To employ nonparametric or machine learning estimators
in the presence of possibly high-dimensional nuisance parameters,
we propose a double debiased machine learning (DML) estimator.
Utilizing a doubly robust moment function
based on a Neyman-type orthogonal score
and cross-fitting, we give high-level conditions under which the nuisance estimators do not affect the first-order large sample distribution of the DML estimators.
Specifically, we give the high-level conditions on the mean-squared convergence rates on the first-step estimators, as for the semiparametric models in  \cite{CCDDHNR}.
The nuisance estimators for the conditional expectation function and the conditional density can be kernel and series estimators, as well as modern ML methods, such as lasso and deep neural networks.
See \cite{CCDDHNR} and \cite{AtheyImbens} for potential ML methods, such as ridge, boosted trees, and various ensembles of these methods.
As each ML method has its strength and weakness depending on the data generating process and applications,
it is desired to flexibly employ various nuisance estimators.
High-dimensional control variables are accommodated via the nuisance estimators; for example, lasso allows the dimension of $X$ to grow with the sample size.


Our paper is related to a growing literature on testing monotonicity in regression problems such as \cite{Bowmanetal1998}, \cite{Ghosaletal2000}, \cite{Gijbelsetal2000}, \cite{HallHeckman2000},
\cite{DumbgenSpokoiny2001}, \cite{Durot2003}, \cite{Beraudetal2005}, \cite{WangMeyer2011}, \cite{Chetverikov2019} and \cite{HsuLiuShi2019}. The main difference between our GMS method and the previously suggested tests is that we rely on a two-step estimation procedure when computing the moments, with the first step consisting of estimating the generalized propensity score, i.e.\ the conditional density of a treatment dose given the covariates, and/or the conditional mean function.  For this reason, it is necessary to take into account the behavior of the first step when we derive the limiting behavior of the estimated moment inequalities underlying our test.


We investigate the finite sample behavior of the proposed test approach in a simulation study and also extend our method to testing conditional monotonicity given observed covariates. As an empirical illustration, we apply our test to data from an experimental study on Job Corps, see \cite{ScBuGl01} and \cite{ScBuMc2008}, a program aimed at increasing the human capital of youths from disadvantaged backgrounds in the U.S. We consider hours in academic and vocational training in the first year of the program as the continuous treatment and investigate its association with several labor market outcomes: weekly earnings in the fourth year, earnings and hours worked per week in quarter 16, and a binary employment indicator four years after assignment. For all outcomes, our test clearly rejects weakly negative monotonicity in the treatment when considering treatment doses between 40 and 3000 hours of training. In contrast, weakly positive monotonicity is not refuted at conventional levels of statistical significance. When, however, splitting the treatment range into 3 brackets of 40 to 1000, 1000 to 2000, and 2000 to 3000 hours, the test points to a violation of weakly negative monotonicity only in the lowest treatment bracket. In the remaining brackets with larger treatment values, we neither reject weakly positive, nor weakly negative monotonicity. Our results are consistent with a concave ADF as for instance found in \cite{Floresetal2012}, suggesting that the marginal effect of training on labor market performance is positive for relatively low treatment doses but decreases as hours in training increase. A potential explanation could be that participants attending more training in the first year might be induced to attain more education in the following years rather than to participate in the labor market.

The paper is organized as follows.  Section \ref{hyp} formulates the hypothesis of weak monotonicity to be tested.
Section \ref{testDML} propose monotonicity tests under DML estimation. Section \ref{sim} presents a Monte-Carlo simulation and discusses how to choose the tuning parameters of the test in practice. Section \ref{app} provides an empirical application to the Job Corps data. Section \ref{cond} adapts the method to testing monotonicity with conditional (rather than unconditional) mean potential outcomes given observed covariates.
Section \ref{conclusion} concludes.
The technical proofs are relegated to the Appendix. An online supplement contains monotonicity tests under nonparametric and parametric estimations of generalized treatment propensity score.

\section{Monotonicity of Continuous Treatment Effect}\label{hyp}
Let $Y(t)$ denote the potential outcome corresponding to the level of treatment intensity
$t\in\mathcal{T}$, where $\mathcal{T}=[a,b]$ with $-\infty< a<b <\infty$.  $Y(t)$ is called the unit-level dose-response function in \cite{HiranoImbens2004}.
Let $\mu(t)=E[Y(t)]$ for $t\in\mathcal{T}$ denote the average of the potential outcome function, also known as the average dose-response function or the average structural function.
In this paper, we are interested in testing if the average dose-response function is weakly increasing in the treatment intensity within a specific range.
We define the null hypothesis of our interest as
\begin{align}
H_{0}:~ \mu(t_1)\geq \mu(t_2),~~\text{for all}~t_{1}\geq t_{2},~\text{for $t_1,t_2\in[t_\ell,t_u]$},\label{eq: null-1}
\end{align}
where $a\leq t_\ell<t_u\leq b$ so that $[t_\ell,t_u]$ is a convex and compact subset of $[a,b]$.  Without loss of generality, we assume that  $[t_\ell,t_u]=[0,1]$.\footnote{If $[t_\ell,t_u]$ is not $[0,1]$,  we can always apply an affine transformation $\phi$ on $t$ so that $\phi(t_\ell)=0$ and $\phi(t_u)=1$. }

Note that the null hypothesis in (\ref{eq: null-1}) has a form that is similar to that in the literature on regression monotonicity, see for instance \cite{HsuLiuShi2019}.  However, the identification of $\mu(t)$ in our case is different from theirs.
We apply the generalized instrumental function approach of \cite{HsuLiuShi2019} and \cite{HsuShen2020} to transform $H_0$ in (\ref{eq: null-1}) to countably many moment inequalities without loss of information.\footnote{The generalized instrumental function approach is a generalization of the instrumental function approach in  \cite{AndrewsShi2013, AndrewsShi2014}.}
To be specific, suppose that $\mu(t)$ is a continuous function on $t=[0,1]$ and $h(t)$ is
a positive weighting function such that $\int_{0}^1h(t)dt<\infty$. Then by Lemma 2.1 of \cite{HsuShen2020}, $H_0$ in (\ref{eq: null-1}) is equivalent to
\begin{align}
&\frac{\int_{t_2}^{t_2+q^{-1}} \mu(s)\cdot h(s) ds}{\int_{t_2}^{t_2+q^{-1}} h(s) ds}
-\frac{\int_{t_1}^{t_1+q^{-1}} \mu(s)\cdot h(s) ds}{\int_{t_1}^{t_1+q^{-1}} h(s) ds}
 \leq 0,~\text{or} \label{eq: inequality-1}\\
&
\int_{t_2}^{t_2+q^{-1}} \mu(s)\cdot h(s) ds
\cdot \int_{t_1}^{t_1+q^{-1}} h(s) ds -\int_{t_1}^{t_1+q^{-1}} \mu(s)\cdot h(s) ds  \cdot
\int_{t_2}^{t_2+q^{-1}} h(s) ds \leq 0 \label{eq: inequality-2}
\end{align}
for any $q=2,\cdots,$ and for any $t_1\geq t_2$ such
that $q\cdot t_1,q\cdot t_2\in \{0,1,2,\cdots,q-1\}$. Equations (\ref{eq: inequality-1}) and (\ref{eq: inequality-2})
hold by the fact that if a function is non-decreasing, then its weighted average over an interval will be
non-decreasing as well when the interval moves to the right.  In addition, by \cite{HsuLiuShi2019}, Equations (\ref{eq: inequality-1}) and (\ref{eq: inequality-2}) contain the same information as the null hypothesis.

%
%
%
%

In the following, we discuss the identification of $\int_{t}^{t+q^{-1}} \mu(s)\cdot h(s) ds$.

\begin{assumption}\label{assu: unconfoundedness}
{\bf (Weak Unconfoundedness):} $Y(t)\perp T \mid X$ for all $t\in \mathcal{T}$.
\end{assumption}

Assumption~\ref{assu: unconfoundedness}  is a commonly invoked identifying assumption based on observational data, also known as conditional independence and selection on observables.
It assumes that conditional on observables $X$, $T$ is as good as randomly assigned, or conditionally exogenous. The observed outcome $Y$ satisfies that $Y=Y(T)$.
We then have the following lemma concerning the identification of $\int_{t}^{t+q^{-1}} \mu(s)\cdot h(s) ds$.
Let $p(t,x)=f_{T|X}(t|x)$ be the generalized propensity score, which is the conditional density of the treatment given the covariates and $p(t,x)>0$ for all $t$ and $x$.

\begin{lemma}\label{Lid}
Suppose Assumption \ref{assu: unconfoundedness} holds. Let $h(t)>0$  for all $t$ be
a known weight function such that $\int_{0}^1h(t)dt<\infty$. Then for $r>0$,
\begin{align*}
&\int_t^{t+r} \mu(s) h(s) ds = E\left[ \frac{Y}{p(T,X)} \cdot h(T)\cdot  1(T\in[t,t+r]) \right].
\end{align*}
\end{lemma}

We now apply Lemma 2.1 of \cite{HsuShen2020} and the identification result in Lemma 2.1 to transform $H_0$ in (\ref{eq: null-1}) to countably many moment inequalities based on which we will construct our test.
For $\ell=(t_1,t_2,q^{-1})\in[0,1]^2\times (0,1]$, define
\begin{align}
&{\cal L}=
\Big\{\ell=(t_1,t_2,q^{-1}):~q\cdot(t_1,t_2)\in\{0,1,2,\cdots,q-1\}^{2},
~t_1> t_2, \text{, and } q=2,3,\cdots\Big\}.\label{eq: Gc-cube}
\end{align}
For each $\ell$, we define
\begin{align*}
\nu_1(\ell)=E\left[\frac{Y}{p(T,X)} 1(T\in[t_1,t_1+q^{-1}])\right],
~~~\nu_2(\ell)=E\left[\frac{Y}{p(T,X)} 1(T\in[t_2,t_2+q^{-1}])\right].
\end{align*}


\begin{lemma}\label{lemma: H0 transformed}
Suppose Assumption \ref{assu: unconfoundedness} holds.  Assume that $\mu(t)$ is continuous in $t$.
Then $H_0$ in (\ref{eq: null-1}) is equivalent to
\begin{align}
H_0':&~\nu(\ell)=\nu_2(\ell)-\nu_1(\ell)\leq 0~\text{for any  $\ell=(t_1,t_2,q^{-1})\in\mathcal{L}$}.\label{eq: null-2}
\end{align}
\end{lemma}

The proof of Lemma \ref{lemma: H0 transformed} is a direct implication of (\ref{eq: inequality-2}) and Lemma \ref{Lid}.  To see this, set $h(t)=1$ and note that $\int_t^{t+r} h(s) ds=r$.  By (\ref{eq: inequality-2}) and Lemma \ref{Lid}, for any $\ell\in \mathcal{L}$,
\begin{align*}
&\int_{t_2}^{t_2+q^{-1}} \mu(s)\cdot h(s) ds
\cdot \int_{t_1}^{t_1+q^{-1}} h(s) ds -\int_{t_1}^{t_1+q^{-1}} \mu(s)\cdot h(s) ds  \cdot
\int_{t_2}^{t_2+q^{-1}} h(s) ds\leq 0\\
\text{iff}&~
\nu_2(\ell) r-\nu_1(\ell) r\leq 0\\
\text{iff}&~\nu(\ell)=\nu_2(\ell)-\nu_1(\ell)\leq 0.
\end{align*}
In Lemma \ref{lemma: H0 transformed}, we pick $h(t)=1$ for simplicity, but the result also holds for any other known  valid wight function $h(t)$.

\section{DML Monotonicity Test}
\label{testDML}
To deliver a reliable distributional approximation in practice,
the double debiased ML (DML) method contains two key ingredients: a doubly robust moment function and cross-fitting.
The doubly robust moment function reduces sensitivity in estimating $\nu(\ell)$ with respect to nuisance parameters.\footnote{
Our estimator is doubly robust in the sense that it consistently estimates $\nu(\ell)$ if either one of the nuisance functions $E[Y|T,X]$ or $f_{T|X}$ is misspecified.
The rapidly growing ML literature has utilized this doubly robust property
to reduce regularization and modeling biases in estimating the nuisance parameters by ML or nonparametric methods;
for example, \cite{BCH14RES}, \cite{Farrell15},  \cite{BCFH17}, \cite{FLM}, \cite{CEINR}, \cite{CCDDHNR}, \cite{RotheFirpo}, and references therein.
}
Cross-fitting removes bias induced by overfitting and achieves stochastic equicontinuity without strong entropy conditions.
Our work builds on the results for semiparametric models in \cite{IchimuraNewey22QE}, \cite{CEINR},  \cite{CCDDHNR}, and the nonparametric models for continuous treatments in \cite{CL}.

We construct the moment function for our DML estimator by the Gateaux derivative limit.
Denote as $\nu(t,r) = \int_t^{t+r} \mu(s) ds$ and $\gamma(t,x) = E[Y|T=t, X=x]$.
Let $f^0$ be the true pdf of $Z = (Y,T,X)$ and $f_Z^h$ be a pdf approaching a point mass at $Z $ as $h \rightarrow 0$.
\cite{CL} derive the Gateaux derivative of $\mu(t)$ with respect to a deviation from the true distribution $f_Z^h - f^0$ to be
\begin{align*}
\gamma(t,X) - \mu(t) + \frac{Y - \gamma(t,X)}{p(t,X)} f_{T}^h(t).
\end{align*}
Since $\nu(t,r)$ is a linear functional of $\mu(t)$, the Gateaux derivative limit of $\nu(t,r)$ is
\begin{align}
&\lim_{h\rightarrow 0} \int_t^{t+r} \left\{ \gamma(s,X) - \mu(s) + \frac{Y - \gamma(s,X)}{p(s,X)} f_{T}^h(s) \right\} ds
\notag \\
&= E\left[ \frac{Y{\bf 1}(T\in [t, t+r])}{p(T,X)}\bigg|X\right] - \nu(t,r) +  \frac{Y - \gamma(T,X)}{p(T,X)}{{\bf 1}(T\in[t, t+r])},
\label{EDR}
\end{align}
and it follows that
\begin{align}
\nu(t,r)=E\left[E\Big[ \frac{Y{\bf 1}(T\in [t, t+r])}{p(T,X)}\bigg|X\Big] +  \frac{Y - \gamma(T,X)}{p(T,X)}{{\bf 1}(T\in[t, t+r])}\right].\label{eq: DML nu iden}
\end{align}

We propose a DML estimator for $\nu(\ell)$ based on (\ref{eq: DML nu iden}):

\begin{changemargin}{0.6cm}{0cm}
\begin{itemize}
\item[Step 1.]
(Cross-fitting) For some fixed $K \in \{2,...,n\}$, a $K$-fold cross-fitting partitions the observation indices into $K$ distinct groups $I_k$, $k=1,..., K$, such that the sample size of each group is the largest integer smaller than $n/K$. Let $n_k$ denote the number of observations in group $I_k$ for $k=1,..., K$.
For $k \in \{1,...,K\}$, the estimators $\hat \gamma_k(t, x)$ and $\hat p_k(t , x)$
for $\gamma(t,x)$ and $p(t,x)$
use observations not in $I_k$ and satisfy Assumption~\ref{A1st} below.

\item[Step 2.]
(Double robustness) The DML estimator is defined as
\begin{align*}
\hat\nu_{DML}(\ell) &= \hat\nu_{2,DML}(\ell)-\hat\nu_{1,DML}(\ell)~\text{where for $j=1$ and 2,}\\
\hat\nu_{j,DML}(\ell)&=
\frac{1}{K}\sum_{k=1}^K \frac{1}{n_k}\sum_{i \in I_k}
\left\{\int_{t_j}^{{t_j}+q^{-1}} \hat \gamma_k(s,X_i) ds + \frac{Y_i - \hat \gamma_k(T_i,X_i)}{\hat p_k(T_i,X_i)}{{\bf 1}(T_i\in[t_j, t_j+q^{-1}])}\right\}
\end{align*}
and $\int_{t_j}^{t_j+q^{-1}} \hat \gamma_k(s,X_i) ds$ is approximated by a numerical integration
$M^{-1}\sum_{m=1}^M  \hat \gamma_k(s_m,X_i) 1(s_m\in [t_j, t_j+q^{-1}])$, with a set of equally spaced grid points $\{s_0 = t_\ell, s_1,..., s_M = t_u\}$ over $[t_\ell, t_u]$.
\end{itemize}
\end{changemargin}

We use $\|\cdot\|_2$ to denote the $L_2$-norm, e.g.
$\|\hat\gamma_k-\gamma\|_2 = \left(\int_\mathcal{X}\int_\mathcal{T}\left( \hat\gamma_k(t,x) - \gamma(t,x) \right)^2 f_{TX}(t, x) dtdx\right)^{1/2}$
and
$\|\hat p_k-p\|_2 = \left(\int_\mathcal{X}\int_\mathcal{T}\left( \hat p_k(t,x) - p(t,x) \right)^2 f_{TX}(t, x) dtdx\right)^{1/2}$.

\begin{assumption}[DML]
For any $k \in\{1,...,K\}$,
\begin{itemize}
\item[(i)] $\|\hat\gamma_k-\gamma\|_2 = o_p(1)$ and
$\|\hat p_k- p\|_2 = o_p(1)$.

\item[(ii)] $\sqrt{n} \|\hat\gamma_k-\gamma\|_2 \|\hat p_k-p\|_2 = o_p(1)$.

\item[(iii)]
The total variation of $\hat \gamma_k$ is finite with probability approaching one.

\item[(iv)]
$p(T, X)$ is bounded away from zero and $var(Y|T,X)$ is bounded above almost surely.

\end{itemize}
\label{A1st}
\end{assumption}

Assumptions~\ref{A1st}(i) and (ii) are the typical conditions on the mean-squared convergence rates, as in \cite{CCDDHNR}.
Assumption~\ref{A1st}(iii) is to control the approximation error of the numerical integration.

\begin{lemma}[DML]
Let Assumptions~\ref{assu: unconfoundedness} and \ref{A1st} hold.
Let $\sqrt{n}/M \rightarrow 0$.
Then uniformly over $\ell\in\mathcal{L}$,
\begin{align}
\sqrt{n}(\hat\nu_{DML}(\ell) - \nu(\ell)) =& n^{-1/2}\sum_{i=1}^n\phi_{\ell,DML}(Y_i,T_i,X_i)+o_p(1)~\text{where}\notag\\
\phi_{\ell,DML}(Y,T,X)=&
E\left[ \frac{Y{\bf 1}(T\in [t_2, t_2+q^{-1}])}{p(T,X)}\bigg|X\right] +  \frac{Y - \gamma(T,X)}{p(T,X)}{{\bf 1}(T\in[t_2, t_2+q^{-1}])} \notag\\
&-E\left[ \frac{Y{\bf 1}(T\in [t_1, t_1+q^{-1}])}{p(T,X)}\bigg|X\right]-  \frac{Y - \gamma(T,X)}{p(T,X)}{{\bf 1}(T\in[t_1, t_1+q^{-1}])}- \nu(\ell). \label{eq: influ DML}
\end{align}
Also, $\sqrt{n}(\hat{\nu}_{DML}(\cdot)-\nu(\cdot))\Rightarrow \Phi_{h_{DML}}(\cdot)$ where
$\Phi_{h_{DML}}(\cdot)$ is a Gaussian process
with variance-covariance kernel $h_{DML}(\ell_1,\ell_2)=E[\phi_{\ell_1,DML}(Y,T,X) \phi_{\ell_2,DML}(Y,T,X)]$.
\label{TIF}
\end{lemma}

Lemma \ref{TIF} establishes the limiting behavior of DML estimators for $\nu$'s.
Let $\hat{\sigma}^2_{\nu,DML}(\ell)={K}^{-1}\sum_{k=1}^K n^{-1}_k\sum_{i \in I_k}   \hat{\phi}^2_{\ell,DML}(Y_i,T_i,X_i)$ where
\begin{align}
\hat{\phi}_{\ell,DML}(Y_i,T_i,X_i)=&\left\{\int_{t_2}^{{t_2}+q^{-1}} \hat \gamma_k(s,X_i) ds + \frac{Y_i - \hat \gamma_k(T_i,X_i)}{\hat p_k(T_i,X_i)}{{\bf 1}(T_i\in[t_2, t_2+q^{-1}])}\right\} \notag\\
&-\left\{\int_{t_1}^{{t_1}+q^{-1}} \hat \gamma_k(s,X_i) ds + \frac{Y_i - \hat \gamma_k(T_i,X_i)}{\hat p_k(T_i,X_i)}{{\bf 1}(T_i\in[t_1, t_1+q^{-1}])}\right\}-\hat{\nu}_{DML}(\ell)
\label{eq: estimated influ DML}
\end{align}
and $\hat{\sigma}^2_{\nu,DML}(\ell)$ will be
a consistent estimator for the asymptotic variance of $\sqrt{n}(\hat{\nu}_{DML}(\ell)-{\nu}(\ell))$ under the assumptions Lemma \ref{TIF}.  Let $\hat{\sigma}_{\nu,\epsilon,DML}(\ell)=\max\{\hat{\sigma}_{\nu,DML}(\ell), \epsilon\cdot \hat{\sigma}_{\nu,DML}(0,1/2,1/2)\}$, by which we manually bound the variance estimator away from zero.
To test the null hypothesis $H_0'$, we make use of a Cram\'{e}r-von Mises test
statistic defined as
\begin{align}
\widehat{T}_{DML}= \sum_{\ell\in\mathcal{L}} \max\Big\{ \sqrt{n}\frac{\hat{\nu}_{DML}(\ell)}{\hat{\sigma}_{\nu,\epsilon,DML}(\ell)},
0 \Big\}^2 Q(\ell),
\label{eq: test statistic}
\end{align}
where $Q$ is a weighting function such that $Q(\ell)>0$ for all  $\ell \in \mathcal{L}$ and
$\sum_{\ell\in\mathcal{L}}Q(\ell)<\infty$.

We next define the simulated critical value for our test.   We first introduce a multiplier bootstrap method
that can simulate a process that converges to the same limit as $\sqrt{n}(\hat{\nu}_{DML}(\ell)-{\nu}(\ell))$.
Let $\{U_i:~ 1\leq i\leq n\}$ be a sequence of i.i.d.\ random variables that satisfy Assumption \ref{assu: U-1}.
We construct the simulated process as
\begin{align} \label{eq: simulated process-mu}
&\widehat{\Phi}_{\nu,DML}^u(\ell)=\frac{1}{\sqrt{n}}\sum_{i=1}^nU_i\cdot\hat{\phi}_{\ell,DML}(Y_i,T_i,X_i),
\end{align}
where $\hat{\phi}_{\ell,np}(Y_i,T_i,X_i)$ is the estimated influence function defined in (\ref{eq: estimated influ DML}).
Under specific regularity conditions, we can show that the simulated process weakly converges to a Gaussian process conditional on the sample path with probability approaching one and that this limiting Gaussian process corresponds to the limiting process of $\sqrt{n}(\hat{\nu}_{np}(\ell)-\nu(\ell))$.

We adopt the GMS method to construct the simulated critical value as
\begin{align*}
\hat{c}^{\eta}_{DML}(\alpha)& =\sup\left\{q\Big|P^u
\left(\sum_{\ell\in \mathcal{ L}} \max\Big\{\frac{\widehat{\Phi}_{\nu,DML}^u(\ell)}{\hat \sigma_{\nu,\epsilon,DML}(\ell)}+\hat{\psi}_{\nu,DML}(\ell),0\Big\}^2 Q(\ell)\leq q\right)\leq 1-\alpha +\eta \right\}+\eta,\\
\hat{\psi}_{\nu,DML}(\ell)&=-B_n \cdot 1\left(\sqrt{n}\cdot \frac{\hat{\nu}_{DML}(\ell)}{\hat \sigma_{\nu,\epsilon,DML}(\ell)} <-a_n\right),
\end{align*}
in which $a_n$ and $B_n$ satisfy Assumption \ref{assu: GMS}.\footnote{The GMS approach is similar to the recentering method of \cite{Hansen2005} and \cite{DonaldHsu2016}, and the contact approach of \cite{Lintonetal2010}.}

The decision rule is then given by
\begin{align}
\text{Reject $H'_0$ if $\widehat{T}_{DML}>\hat{c}^{\eta}_{DML}(\alpha)$. } \label{eq: decision rule}
\end{align}

\begin{assumption}\label{assu: U-1}
	$\{U_i:~ 1\leq i\leq n\}$ is a sequence of i.i.d.\ random
	variables that is independent of the sample path of $\{(Y_i,X_i,
	T_i):~ 1\leq i\leq n\}$ such that $E[U_i]=0$, $E[U_i^2]=1$, and
	$E[|U_i|^{2+\delta}]<C$ for some $\delta>0$ and $C>0$.
\end{assumption}

\begin{assumption}\label{assu: GMS}
	(i)
	$a_n$ is a sequence of non-negative numbers satisfying
	$\lim_{n\rightarrow\infty}a_n=\infty$ and
	$\lim_{n\rightarrow\infty}a_n/\sqrt{n}=0$.\\
	(ii) $B_n$ is a sequence of non-negative numbers satisfying
	that $B_n$ is non-decreasing,
	$\lim_{n\rightarrow\infty}B_n=\infty$ and
	$\lim_{n\rightarrow\infty}B_n/a_n=0$.
	
\end{assumption}

\begin{thm}\label{thm: size and power}
	Suppose that Assumptions \ref{assu: unconfoundedness}, \ref{A1st}, \ref{assu: U-1} and \ref{assu: GMS} hold.  Then the following statements are true:
	\begin{enumerate}[(a)]
		\item
		Under $H_0$, $\lim_{n\rightarrow\infty} P(\widehat{T}_{DML}>\hat{c}^{\eta}_{DML}(\alpha))\leq \alpha$;
		\item
		Under $H_1$, $\lim_{n\rightarrow\infty} P(\widehat{T}_{DML}>\hat{c}^{\eta}_{DML}(\alpha))=1$.
	\end{enumerate}
\end{thm}


%

\medskip

The high-level conditions in Assumption~\ref{A1st} are attainable by various estimators, in particular, kernel, series, deep neural networks, and lasso.
The theory of the conventional nonparametric kernel and series methods is well established.
Recently \cite{FLM} provide $\|\hat\gamma_k - \gamma\|_2$ of deep neural networks.
\cite{CL} propose GPS estimators that utilize generic estimators of the conditional mean function.
Specifically, Lemmas 1 and 2 in \cite{CL} provide the convergence rates for their GPS estimators using the deep neural networks in \cite{FLM}, i.e. $\|\hat p_k-p\|_2$.\footnote{
The convergence rates in Lammas 1 and 2 of \cite{CL} can be shown to hold uniformly over $t\in\mathcal{T}$, so we can obtain $\|\hat p_k-p\|_2$.
The additional assumption for the MultiGPS estimator in Lemma 2 \cite{CL} is  $\sup_{t\in\mathcal{T}}\left\| \hat\mu\left(h_1^{d_t}g_{h_1}(T-t); X\right) - {\mathbb{E}}[h_1^{d_t}g_{h_1}(T-t)|X] \right\|_{F_X} = O_p(R_{1n})$.
Then Lemma 3 in \cite{CL} provides the sufficient conditions.
}
So Assumptions~\ref{A1st}(i) and (ii) are attainable by the deep neural networks in \cite{FLM} and \cite{CL}.
In the rest of this section, we provide the sufficient low-level conditions for lasso methods.

\subsection{Step 1 lasso}
We illustrate how to employ lasso methods to estimate the nuisance conditional mean function $\gamma(t,x)$ and the generalized propensity score $f_{T|X}(t|x)$.
We provide sufficient conditions to verify the high-level Assumption~\ref{A1st}.
We modify the penalized local least squares estimator of $\gamma(t, X)$ in \cite*{SUZ} (SUZ, hereafter).
We use the conditional density estimator in SUZ.
For completeness, we present the estimators and asymptotic theory in SUZ and refer readers to SUZ for details.

Let $b(T, X)$ be a $p\times 1$ vector of basis functions.
We approximate $\gamma(t,x)$ by $b(t,x)'\theta$.
The lasso estimator $\hat\gamma_k(t,x) = b(t,x)'\hat\theta_k$ for $k\in\{1,...,K\}$, where
\begin{align}
\hat\theta_k = \arg\min_\theta \frac{1}{2(n-n_{k})}\sum_{i\notin I_k} (Y_i - b(T_i, X_i)'\theta)^2 + \frac{\lambda}{n-n_{k}} \|\hat \Xi_k \theta\|_1,
\label{SUZ3.4}
\end{align}
where $n_{k} =  \sum_{i=1}^n {\bf 1}\{i \in I_k\}$,
$\|\cdot\|_1$ denotes the $L_1$ norm,
$\lambda = \ell_n(\log(p\vee n)n)^{1/2}$ for some slowly diverging sequence $\ell_n$, and
$\hat\Xi_k = diag(\tilde l_{k1},...,\tilde l_{kp})$ is a generic penalty loading matrix computed by Algorithm~\ref{ASUZ3.1} below
from the iterative Algorithm 3.1 in SUZ.
Denote as $\| f(X) \|_{\mathbb{P}_{nk},2} = \big((n-n_k)^{-1}\sum_{i \notin I_k} f(X_i)^2\big)^{-1/2}$ for a generic function $f(\cdot)$.

\begin{algorithm}[SUZ Algorithm 3.1]
For $k\in\{1,...,K\}$,
\begin{enumerate}
\item
Let $\hat\Xi^0_k = diag(\tilde l_{k1}^0,...,\tilde l_{kp}^0)$, where $\tilde l_{kj}^0 = \| Yb_j(T, X)\|_{\mathbb{P}_{nk}, 2}$.
Compute $\hat\theta^0_k$ by (\ref{SUZ3.4}) with $\hat\Xi^0_k$ in place of $\hat\Xi_k$.
Let $\hat\gamma^0_k(T_i, X_i) = b(T_i, X_i)'\hat\theta^0_k$.

\item For $s=1,...,S$ for some fixed positive integer $S$,
compute $\hat\Xi^s_k = diag(\tilde l_{k1}^s,...,\tilde l_{kp}^s)$, where $\tilde l_{kj}^s = \| (Y - \hat \gamma_k^{s-1}(T,X))b_j(T, X)\|_{\mathbb{P}_{nk}, 2}$.
Compute $\hat\theta^s_k$ by (\ref{SUZ3.4}) with $\hat\Xi^s_k$ in place of $\hat\Xi_k$.
and $\hat\gamma^s_k(T_i, X_i) = b(T_i, X_i)'\hat\theta^s_k$.
\end{enumerate}
\label{ASUZ3.1}
\end{algorithm}

Let the final penalty loading matrix $\hat\Xi_k=\hat\Xi^S_k$ from Algorithm~\ref{ASUZ3.1}.
Then the lasso estimator $\hat\gamma_k(t,x) = b(t,x)'\hat\theta_k$ for $k\in\{1,...,K\}$ from (\ref{SUZ3.4}).

\bigskip

To estimate the conditional density $p(t,x) = f_{T|X}(t|x)$, first estimate the conditional CDF $F_{T|X}$ by the logistic distributional lasso regression and then take the numerical derivative.
Let $b(X)$ be a $p\times 1$ vector of basis functions.
We approximate $F_{T|X}(t|x)$ by $\Lambda(b(x)'\beta_t)$, where $\Lambda$ is the logistic CDF.
For $k\in\{1,...,K\}$,
$\hat F_{T|X_k}(t|x) = \Lambda(b(X)'\hat\beta_{tk})$, where
\begin{align}
\hat\beta_{tk} = \arg\min_\beta \frac{1}{ n-n_k}\sum_{i\notin I_k} M({\bf 1}\{T_i \leq t\}, X_i; \beta) + \frac{\tilde \lambda}{n-n_k}\|\hat\Psi_{tk}\beta\|_1
\label{SUZ3.6}
\end{align}
where $n_{k} =  \sum_{i=1}^n {\bf 1}\{i \in I_k\}$,
$M(y, x; g) = -[y\log(\Lambda(b(x)'g)) + (1-y)\log(1-\Lambda(b(x)'g))]$ is the logistic likelihood,
the penalty $\tilde\lambda = 1.1\Phi^{-1}(1-r/\{p\vee nh_1\}) n^{1/2}$,
for some $r\rightarrow 0$ and $h_1\rightarrow 0$, and $\Phi$ is the standard normal CDF.
A generic penalty loading matrix $\hat\Psi_{tk}$ is computed by Algorithm~\ref{ASUZ3.2} below
from the iterative Algorithm 3.2 in SUZ.

\begin{algorithm}[SUZ Algorithm 3.2]
For $k\in\{1,...,K\}$,
\begin{enumerate}
\item
Let $\hat \Psi_{tk}^0 = diag(l_{tk,1}^0,...,l_{tk,p}^0)$, where $l_{tk,j}^0 = \|{\bf 1}\{T\leq t\} b_j(X)\|_{\mathbb{P}_{nk},2}$.
Compute $\hat\beta_{tk}^0$  by (\ref{SUZ3.6})
with $\hat \Psi_{tk}^0$ in place of  $\hat \Psi_{tk}$
and $\hat F_{T|X_k}^0(t|x) = \Lambda(b(x)'\hat\beta_{tk}^0)$.

\item For $s=1,...,S$, compute $\hat\Psi_{tk}^s = diag(l_{tk,1}^s,...,l_{tk,p}^s)$, where
$l_{tk,j}^s = \Big\|\Big({\bf 1}\{T\leq t\}  - \hat F_{T|X_k}^{s-1}(t|X)\Big)b_j(X)\Big\|_{\mathbb{P}_{nk},2}$.
Compute $\hat\beta_{tk}^s$  by (\ref{SUZ3.6})
with $\hat \Psi_{tk}^s$ in place of  $\hat \Psi_{tk}$
and $\hat F_{T|X_k}^s(t,x) = \Lambda(b(x)'\hat\beta_{tk}^s)$.

\end{enumerate}
\label{ASUZ3.2}
\end{algorithm}

Let the final penalty loading matrix $\hat \Psi_{tk} = \hat\Psi_{tk}^S$ from Algorithm~\ref{ASUZ3.2}.
Compute $\hat F_{T|X_k}(t|x) = \Lambda(b(X)'\hat\beta_{tk})$ from (\ref{SUZ3.6}).
Then the conditional density estimator
\[
\hat p_k(t,x) = \frac{\hat F_{T|X_k}(t+h_1|x) - \hat F_{T|X_k}(t-h_1|x)}{2h_1}.
\]

Assumption~\ref{ASUZ} collects the conditions in Theorems 3.1 and 3.2 in SUZ.
Following SUZ's notations, denote as $\|\cdot\|_{\mathbb{Q},q}$ the $L^q$ norm under measure $\mathcal{Q}$
and $\mathbb{P}$ assigns probability $1/n$ to each observation.
\begin{assumption}[Lasso]
Let $\mathcal{T}$ be a compact subset of the support of $T$ and
$\mathcal{X}$ be the support of $X$.

\begin{enumerate}
\item
\begin{enumerate}
\item $\|\max_{j \leq p} |b_j(T, X)|\|_{\mathbb{P}, \infty} \leq \zeta_n$
and
$\underline{C} \leq \mathbb{E}\left[b_j(T, X)^2\right] \leq 1/\underline{C}$, for some positive constant $\underline{C}$, $j=1,...,p$.

\item $\sup_{t \in \mathcal{T}} \max(\|\beta_t\|_0, \|\theta\|_0) \leq s$ for some $s$ which possibly depends on $n$,
where $\|\theta\|_0$ denotes the number of nonzero coordinates of $\theta$.


\item
For the approximation error,
$\sup_{t\in\mathcal{T}} \|F_{T|X}(t|X) - \Lambda(b(X)'\beta_t)\|_{\mathbb{P}, \infty} = O_p((s^2\zeta_n^2\log(p \vee n)/n)^{1/2})$
and
$\left\|\gamma(T,X) - b(T,X)'\theta\right\|_{\mathbb{P},\infty}= o_p\left(\left(s^2\zeta_n^2\log(p\vee n)/n\right)^{1/2}\right)$.

\item
$p(t,x)$ is second-order differentiable w.r.t. $t$ with bounded derivatives uniformly over $(t,x)\in\mathcal{T}\times\mathcal{X}$.

\item
$\zeta_n^2s^2\ell_n^2\log(p\vee n)/(nh_1) \rightarrow 0$, $nh_1^5/(\log(p\vee n))\rightarrow 0$.
\end{enumerate}

\item
\begin{enumerate}
\item There exists some positive constant $\underline{C} < 1$ such that $\underline{C}  \leq p(t,x) \leq 1/\underline{C}$ uniformly over $(t,x)\in\mathcal{T}\times\mathcal{X}$.

\item
$\gamma(t,x)$ is three times differentiable with all three derivatives being bounded uniformly over $(t,x)\in\mathcal{T}\times\mathcal{X}$.


\end{enumerate}

\item
There exists a sequence $\ell_n\rightarrow \infty$ such that, with probability approaching one, $0 < \kappa' \leq
\inf_{\delta\neq 0, \|\delta\|_0 \leq s\ell_n} \frac{\|b(T,X)'\delta\|_{\mathbb{P}_n,2}}{\|\delta\|_2}
\leq
\sup_{\delta\neq 0, \|\delta\|_0 \leq s\ell_n} \frac{\|b(T,X)'\delta\|_{\mathbb{P}_n,2}}{\|\delta\|_2}
\leq \kappa'' < \infty.$


\end{enumerate}

\label{ASUZ}
\end{assumption}

Let Assumption~\ref{ASUZ} hold.
Then Theorems 3.1 and 3.2 in SUZ imply that
$\sup_{(t, x)\in\mathcal{T}\times\mathcal{X}}|\hat\gamma_k(t, x) - \gamma(t, x) |= O_p(A_n)$,
where $A_n = \ell_n(\log(p\vee n) s^2\zeta_n^2/n)^{1/2}$
and
$\sup_{(t, x)\in\mathcal{T}\times\mathcal{X}}|\hat p_k(t, x) - p(t, x)| = O_p(B_n)$, where $B_n = h_1^{-1}(\log(p\vee n) s^2\zeta_n^2/n)^{1/2}$.
Then we can obtain the same rates for the root-mean-squared rates $\|\hat\gamma_k - \gamma\|_2 $
and $\|\hat p_k - p\|_2$ to verify Assumption~\ref{A1st}.
Therefore a sufficient condition of Assumption~\ref{ASUZ}(i) is $A_n  \rightarrow 0$ and $B_n \rightarrow 0$.
And a sufficient condition of Assumption~\ref{ASUZ}(ii) is $\sqrt{n}A_n B_n \rightarrow 0$.

\section{Simulation}\label{sim}

This section provides a simulation study to examine the finite sample performance of the proposed test. To implement our test in practice, one has to choose several tuning parameters in advance. We make the following propositions concerning the choice of these parameters and present related Monte Carlo simulation results further below.

\begin{enumerate}
	
	\item
	Instrumental functions: We opt for using a set of indicator functions of countable hypercubes. For $\ell=(t_1,t_2,q^{-1})\in[0,1]^2\times (0,1]$, define
	\begin{align}
	&{\cal L}=
	\Big\{\ell=(t_1,t_2,q^{-1}):q\cdot(t_1,t_2)\in\{0,1,2,\cdots,q-1\}^{2},\nonumber\\
	&~~~~~~~~~~~~~~~~t_1> t_2,\text{ and } q=2,\cdots,q_1\Big\},
	\end{align}
	where $q_1$ is a natural number and is chosen such that the expected sample size of the smallest cube is around 50. Our simulations show that the results are robust to various expected sample sizes.
	
	\item
	{\bf $Q(\ell)$:}  The distribution $Q(\ell)$ assigns weight $\propto
	q^{-2}$ to each $q$ and for each $q$, $Q(\ell)$ assigns an equal weight to each instrumental function with last element of $\ell$ equal to $q^{-1}$. Recall that for each $q$, there are $(q(q+1)/2)$ instrumental functions with the last element of $\ell$ equal to $q^{-1}$.
	
	\item
	{\bf $a_n$, $B_n$, $\epsilon$, $\eta$:} We set $a_{n}=0.15\cdot\ln(n)$, $B_{n}=0.85\cdot\ln(n)/\ln\ln(n)$, $\epsilon=10^{-6}$, and $\eta=10^{-6}$ as suggested by \cite{HsuLiuShi2019}. These choices are used in all the simulations that we report below and seem to perform well.
	
\end{enumerate}

For all data generating processes (DGPs), the continuous treatment variable $T$, the control variables $X$, and the error term $U_y$ are generated as follows
\begin{align*}
&T=(3.6+X'\beta)/7.2+0.5 U_t,\\
&X=(X_{1},\dots,X_{100})'\sim\mathcal{N}(0,\Sigma),\\
&U_y \sim \mathcal{N}(0,1),
\end{align*}
where the $(i,j)$-entry $\Sigma_{ij}=(0.5)^{|i-j|}$ for $i,j=1,\dots,100$,  $U_t\sim\mathcal{N}(0,1)$, and $U_y$, $U_t$, and $X$ are mutually independent. We set $\beta_j=1/j^2$ for mild dependence between $X_j$ and $\beta_j=1/j$ for strong dependence between $X_j$. Three cases of the potential outcomes are studied:
\begin{itemize}
	\item[DGP 1:] $Y=U_y$,
	\item[DGP 2:] $Y= X'\beta T+T^2+X'\beta+U_y$,
	\item[DGP 3:] $Y= X'\beta T+\sin(\pi T)+X'\beta+U_y$.
\end{itemize}

In DGP 1, $\mu(t)=0$, and $H_0$ holds with moment equalities. In this case, we expect that the size of the proposed test will achieve the nominal level since every moment would hold with equality. In DGP 2, $\mu(t)= t^2$, and $H_0$ holds with strict moment inequalities. In this case, we expect the size will converge to zero since every moment would hold with strict inequality. This is because the test statistics will converge to zero and the critical value is bounded away from zero. In DGP 3, $\mu(t)=\sin(\pi T)$, and $H_0$ does not hold. In this case, we expect the power will increase with the sample size.

In these DGPs, $1+d_x=101$. We consider samples of sizes $n = 200$, $400$, $800$, and $1600$. For $q_1$, we set $q_1=4$ for $n=200$, $q_1=8$ for $n=400$, $q_1=16$ for $n=800$, and $q_1=32$ for $n=1600$. The number of subsamples used for cross-fitting is $K\in\{2,5,10\}$. All our Monte Carlo results are based on $1000$ simulations. In each simulation, the critical value is approximated by $1000$ bootstrap replications. The nominal size of the test is set at $10\%$.

To estimate the conditional mean function $\gamma(t,x) = E[Y|T=t, X=x]$, we employ the lasso regression, where the penalization parameter is chosen via grid search utilizing 10-fold cross validation.  To estimate the conditional density estimation $p(t,X)$, we first estimate $F_{T|X}(t|x)$ by the logistic distributional lasso regression, and then take the numerical derivative. The penalization parameter of the distributional lasso regression is estimated by Algorithm 3.2 of \cite{SUZ}. Also, all lasso estimations include an intercept and the covariates.
For numerical integration in Step 2, we set $M=[n^{2/3}]$, where $[\cdot]$ is the nearest integer. Our test is based on the trimmed generalized propensity score estimator, defined as $\tilde{p}(T_i, X_i)=\max\{\hat{p}(T_i, X_i),0.025\}$, implying that conditional treatment densities below 2.5\% are set to 2.5\%.\footnote{In general, one can follow \cite{DHL2014} and \cite{HLL2020} and trim the estimated generalized propensity scores to prevent them from being too close zero, in order to obtain a more stable IPW estimator whose variance is not affected by extremely low scores.}$^,$\footnote{Based on this trimming rule, around 0.5\% of the samples are trimmed.}

\begin{table}[htp]
	\begin{center}
		\caption{Rejection probabilities of our test for $N=50$}\label{Table: N50}
		\begin{tabular}{cccccccc}
			\hline\hline
			&&\multicolumn{3}{c}{$\beta_j=1/j^2$}&\multicolumn{3}{c}{$\beta_j=1/j$}\\
			\cmidrule(lr){3-5} \cmidrule(lr){6-8}
			DGP&n&K=2&K=5&K=10&K=2&K=5&K=10\\\hline
			1&200&0.121&0.107&0.117&0.118&0.116&0.119\\
			1&400&0.093&0.099&0.100&0.098&0.112&0.118\\
			1&800&0.102&0.120&0.111&0.109&0.114&0.094\\
			1&1600&0.109&0.097&0.110&0.095&0.106&0.086\\
			\hline
			2&200&0.001&0.002&0.000&0.003&0.000&0.000\\
			2&400&0.000&0.000&0.000&0.000&0.002&0.000\\
			2&800&0.000&0.000&0.000&0.000&0.000&0.000\\
			2&1600&0.000&0.000&0.000&0.000&0.000&0.000\\
			\hline
			3&200&0.182&0.207&0.229&0.066&0.088&0.092\\
			3&400&0.423&0.517&0.495&0.143&0.222&0.213\\
			3&800&0.870&0.906&0.911&0.454&0.539&0.550\\
			3&1600&0.999&1.000&1.000&0.898&0.921&0.937\\
			\hline\hline
		\end{tabular}
	\end{center}
\end{table}

Table \ref{Table: N50} shows the rejection probabilities of our test for DGPs 1-3, and the results are consistent with our theoretical findings. For the mild dependence case, the proposed test controls size well in DGP 1 and DGP 2, and the rejection probabilities increase with the sample size and are greater than the nominal size $0.1$ in DGP 3. For the strong dependence case, our test still control size will in both DGP 1 and DGP 2. The power increases with the sample size in DGP 3, but the rejection probabilities are a bit less than the nominal size $0.1$ for $n=200$. Overall, we do not find significant difference for different choices of $K$.

\begin{table}[htp]
	\begin{center}
		\caption{Rejection probabilities of our test for $K=5$ and different $N$}\label{Table: Ns}
		\begin{tabular}{cccccccccc}
			\hline\hline
			&&\multicolumn{4}{c}{$\beta_j=1/j^2$}&\multicolumn{4}{c}{$\beta_j=1/j$}\\
			\cmidrule(lr){3-6} \cmidrule(lr){7-10}
			DGP&n&N=33&N=40&N=50&N=66&N=33&N=40&N=50&N=66\\\hline
			1&200&0.133&0.100&0.107&0.113&0.101&0.121&0.116&0.111\\
			1&400&0.118&0.116&0.099&0.099&0.127&0.114&0.112&0.110\\
			1&800&0.136&0.103&0.120&0.091&0.122&0.108&0.114&0.109\\
			1&1600&0.114&0.101&0.097&0.115&0.100&0.127&0.106&0.130\\
			\hline
			2&200&0.003&0.000&0.002&0.000&0.002&0.000&0.000&0.001\\
			2&400&0.000&0.000&0.000&0.000&0.000&0.001&0.002&0.000\\
			2&800&0.000&0.000&0.000&0.000&0.000&0.000&0.000&0.000\\
			2&1600&0.000&0.000&0.000&0.000&0.000&0.000&0.000&0.000\\
			\hline
			3&200&0.210&0.247&0.207&0.170&0.100&0.072&0.088&0.084\\
			3&400&0.521&0.503&0.517&0.465&0.220&0.221&0.222&0.202\\
			3&800&0.911&0.912&0.906&0.895&0.549&0.554&0.539&0.547\\
			3&1600&0.999&0.999&1.000&1.000&0.933&0.932&0.921&0.933\\
			\hline\hline
		\end{tabular}
	\end{center}
\end{table}

We next investigate the robustness of the performance of our test to the choice of $q_{1}$. Let $N$ denote the expected sample size of the smallest cube. We consider three alternative choices of $q_{1}$, each resulting in $N = 33$, $40$, and $66$, respectively. Table \ref{Table: Ns} shows the rejection probabilities of our test for different choices of $N$. The results suggest that the choice of $q_{1}$ does not affect the test performance much. Therefore, the finite sample behavior of our test appears to be reasonably robust to different values of $q_{1}$.

\section{Empirical application}\label{app}

As an empirical illustration, we apply our test to data from the Job Corps study. The latter was conducted between November 1994 and February 1996 to evaluate the publicly funded U.S.\ Job Corps program and used an experimental design that randomly assigned access to the program. Job Corps targets youths from low-income households who are between 16 and 24 years old and legally reside in the U.S. Program participants obtained on average roughly 1200 hours of vocational and/or academic classroom training as well as housing and board over an average duration of 8 months. We refer to \cite{ScBuGl01} and \cite{ScBuMc2008} for a detailed discussion of the study design and the average effects of program assignment on a range of different outcomes. Their results suggest that Job Corps raises educational attainment, reduces criminal activity, and increases labor market performance measured by employment and earnings, at least for some years after the program.

Particularly relevant for our context is the study by \cite{Floresetal2012}, who consider the length of exposure to academic and/or vocational training as continuously distributed treatment to assess its effect on earnings based on regression and weighting estimators (using the inverse of the conditional treatment density as weight).  As the length of treatment exposure is (in contrast to Job Corps assignment) not random, they impose a selection-on-observables assumption and control for baseline characteristics at Job Corps assignment. While the authors find overall positive average effects of increasing hours in academic and vocational instruction, the marginal effects appear to decrease with length of exposure, pointing to a potential concavity in the association of earnings and time of instruction. Relatedly, \cite{Lee2018} and \cite{CL} assess the effect of hours in training on the proportion of weeks employed in the second year after program assignment based on kernel regression and double machine learning, respectively. Also for this outcome, the plotted regression lines in either study point to a concave association with the treatment dose.\footnote{See also \cite{Huberetal2020}, who use a causal mediation approach to assess the direct effect of the treatment dose on the number of arrests in the fourth year after program assignment when controlling for employment behavior in the second year based on inverse probability weighting and find a non-linear association.}

However, in the light of estimation uncertainty, mere eye-balling of the outcome-treatment associations in empirical applications does not tell us whether specific shape restrictions can be refuted. For this reason, we use our DML method with lasso regression for nuisance parameter estimation to formally test whether weak positive and negative monotonicity can be rejected in the Job Corps data when considering several labor market outcomes. To this end, we define the treatment variable $T$ as the total hours spent in academic and vocational training in the 12 months following the program assignment. Our outcomes $Y$ include
weekly earnings in the fourth year, earnings  and hours worked per week in quarter 16, and a binary employment indicator four years after assignment (i.e.\ in week 208).

For invoking weak unconfoundedness (Assumption 2.1), we consider the same set of pre-treatment covariates $X$ as \cite{Lee2018}, \cite{CL}, and \cite{Huberetal2020}, which overlaps with the control variables of \cite{Floresetal2012}.\footnote{A control variable in \cite{Floresetal2012} we do not have access to is the local unemployment rate which was constructed by matching county-level unemployment rates to individual postal codes of residence, which are only available in a restricted-use data set.} We condition on individual characteristics like age, gender, ethnicity, language competency, education, marital status, household size and income, previous receipt of social aid,  family background (e.g. parents' education), criminal activity, as well as health and health-related behavior (e.g.\ smoking, alcohol, or drug consumption). Conditioning on such a rich set of socio-economic variables appears important, as the satisfaction of weak unconfoundedness relies on successfully controlling for all factors jointly affecting treatment duration and labor market behavior. Furthermore, we include variables that might be associated with the duration in training, namely expectations about Job Corps and interaction with the recruiters, which might serve as proxies for unobserved personality traits (like motivation) that could also affect the outcomes. Finally, we control for pre-treatment outcomes, namely previous labor market participation and earnings, to tackle any confounders that affect the outcomes of interest through their respective pre-treatment values.

The original Job Corps data set consists of $15,386$ individuals prior to program assignment, but a substantial share never enrolled in the program and dropped out of the study, such that there are only $11,313$ individuals with completed follow-up interviews
four years after randomization. Among those, $6,828$ had been randomized into Job Corps and had thus access to academic or vocational training. To define our final evaluation sample, we follow \cite{Floresetal2012}, \cite{Lee2018}, \cite{CL}, and \cite{Huberetal2020} and consider observations with at least 40 hours (or one working week) of training for our analysis, all in all $4,166$ individuals. Among these, there are cases of item non-response in various elements of $X$ measured at the baseline survey, for which we account by the inclusion of missing dummies as additional regressors, while observations with missing values in the outcome of interest need to be dropped when running the respective test. Table \ref{des} provides descriptive statistics for selected covariates $X$ (see \cite{Huberetal2020} for a full list of control variabes) as well as for the treatment $T$ and all outcomes $Y$, including the respective number of nonmissing observations (nonmissing).

\begin{table}[htp]
	\begin{center}
		\caption{Descriptives for selected covariates, treatment, and outcomes}\label{des}
{\small		\begin{tabular}{cccccc}
			\hline\hline
			variable & mean & median & minimum & maximum & nonmissing \\
			\hline
			female & 0.432 & 0.495 & 0.000 & 1.000 & 4166\\
			age & 18.325 & 2.142 & 16.000 & 24.000 & 4166  \\
			white & 0.249 & 0.433 & 0.000 & 1.000 & 4166  \\
			black & 0.502 & 0.500 & 0.000 & 1.000 & 4166  \\
			Hispanic & 0.172 & 0.378 & 0.000 & 1.000 & 4166  \\
			years of education & 10.045 & 1.535 & 0.000 & 20.000 & 4102  \\
			married & 0.016 & 0.126 & 0.000 & 1.000 & 4166  \\
			has children & 0.178 & 0.382 & 0.000 & 1.000 & 4166  \\
			ever worked & 0.145 & 0.352 & 0.000 & 1.000 & 4166  \\
			mean gross weekly earnings & 19.429 & 97.749 & 0.000 & 2000.000 & 4166  \\
			household size & 3.536 & 2.006 & 0.000 & 15.000 & 4101  \\
			mum's years of education & 11.504 & 2.599 & 0.000 & 20.000 & 3397  \\
			dad's years of education  & 11.459 & 2.900 & 0.000 & 20.000 & 2604  \\
			welfare receipt during childhood & 2.064 & 1.189 & 1.000 & 4.000 & 3871  \\
			poor or fair general health & 0.124 & 0.330 & 0.000 & 1.000 & 4166  \\
			physical or emotional problems & 0.043 & 0.203 & 0.000 & 1.000 & 4166  \\
			extent of marijuana use   & 2.540 & 1.549 & 0.000 & 4.000 & 1534  \\
			extent of smoking  & 1.526 & 0.971 & 0.000 & 4.000 & 2171  \\
			extent of alcohol consumption   & 3.140 & 1.210 & 0.000 & 4.000 & 2383  \\
			ever arrested & 0.241 & 0.428 & 0.000 & 1.000 & 4166  \\
			recruiter support    & 1.592 & 1.059 & 1.000 & 5.000 & 4068  \\
			idea about desired training & 0.839 & 0.368 & 0.000 & 1.000 & 4166  \\
			expected months in Job Corps & 6.622 & 9.794 & 0.000 & 36.000 & 4166  \\
			hours in   training ($T$)& 1192.130 & 966.945 & 0.857 & 6188.571 & 4166  \\
			weekly earnings in fourth year ($Y$) & 215.521 & 202.619 & 0.000 & 1879.172 & 4024  \\
			weekly earnings in quarter 16 ($Y$) & 220.933 & 223.078 & 0.000 & 1970.445 & 4015  \\
			weekly hours worked quarter 16 ($Y$) & 28.187 & 22.746 & 0.000 & 84.000 & 4102  \\
			employed in week 208 ($Y$) & 0.627 & 0.484 & 0.000 & 1.000 & 4007  \\
			\hline
		\end{tabular}}
	\end{center}
\end{table}

The choices of nuisance parameters are the same as in the simulations (see the previous section). The number of subsamples used for cross-fitting is $5$, and the expected sample size of the smallest cube is either $40$ or $50$. The lasso estimations include an intercept, the covariates and the squared terms of any non-binary covariates.
The $p$-values of the tests for the various outcomes are calculated based on 1000 bootstrap replications.\footnote{In our empirical study, we do not get unstable IPW $\nu(\ell)$ estimates, so we decide not to apply the trimming method.  Also, we note that 
all estimated generalized propensity scores are greater than $0.0001$ in our empirical study.}

In a first step, we apply the test to a treatment interval of $T \in [40,3000]$, where choosing 3000 hours of training as upper bound of the analysis is motivated by the quickly decreasing number of observations beyond that point.

\begin{table}[htp]
	\begin{center}
		\caption{Test statistic and p-value, $40\leq T \leq 3000$}\label{Table:40_3000}
		\begin{tabular}{ccccccccc}
			\hline\hline
			&\multicolumn{4}{c}{N=40, $t_1>t_2$}&\multicolumn{4}{c}{N=50, $t_1>t_2$}\\
			\cmidrule(lr){2-5} \cmidrule(lr){6-9}
			$H_0:$ &\multicolumn{2}{c}{$\mu(t_1)\geq\mu(t_2)$}& \multicolumn{2}{c}{$\mu(t_1)\leq\mu(t_2)$}
			&\multicolumn{2}{c}{$\mu(t_1)\geq\mu(t_2)$}& \multicolumn{2}{c}{$\mu(t_1)\leq\mu(t_2)$} \\
			$Y$& stat & p-value & stat & p-value& stat & p-value & stat & p-value\\
			\hline
            earny4&0.001&1.000&7.205&0.000&0.001&1.000&6.078&0.000\\
earnq16&0.001&1.000&8.740&0.000&0.001&1.000&8.435&0.000\\
hrswq16&0.001&1.000&9.985&0.000&0.001&1.000&9.613&0.000\\
work208&0.001&0.997&10.397&0.000&0.001&0.998&9.478&0.000\\
			\hline\hline
		\end{tabular}
	\end{center}
		\par
		Note: Outcomes `earny4', `earnq16', `hrswq16', and `work208' are weekly earnings in the fourth year, earnings and hours worked per week in quarter 16, and a binary employment indicator four years after assignment (i.e.\ in week 208). `stat' denotes the test statistic.
	\end{table}

Table \ref{Table:40_3000} reports the test statistics and p-values for all outcomes under both null hypotheses of weakly increasing mean potential outcomes in the treatment ($\mu(t_1)\geq\mu(t_2)$ for $t_1>t_2$) and weakly decreasing mean potential outcomes ($\mu(t_1)\leq\mu(t_2)$), respectively. Our tests clearly reject the latter hypothesis of weakly negative monotonicity for any labor market outcome at the 1\% level of statistical significance. In contrast, weak positive monotonicity is never rejected, as any test yields p-values close to or equal to 1 (or 100\%). Our findings therefore suggest that an increase in the treatment does either increase or at least not reduce the outcome over the treatment range $T \in [40,3000]$.

	\begin{table}[htp]
	\begin{center}
		\caption{Test statistic and p-value, $40\leq T \leq 1000$}\label{Table:40_1000}
		\begin{tabular}{ccccccccc}
			\hline\hline
			&\multicolumn{4}{c}{N=40, $t_1>t_2$}&\multicolumn{4}{c}{N=50, $t_1>t_2$}\\
			\cmidrule(lr){2-5} \cmidrule(lr){6-9}
			$H_0:$ &\multicolumn{2}{c}{$\mu(t_1)\geq\mu(t_2)$}& \multicolumn{2}{c}{$\mu(t_1)\leq\mu(t_2)$}
			&\multicolumn{2}{c}{$\mu(t_1)\geq\mu(t_2)$}& \multicolumn{2}{c}{$\mu(t_1)\leq\mu(t_2)$} \\
			$Y$& stat & p-value & stat & p-value& stat & p-value & stat & p-value\\
			\hline
             earny4&0.004&0.750&11.402&0.000&0.004&0.750&11.562&0.000\\
earnq16&0.017&0.535&5.556&0.000&0.016&0.540&5.427&0.000\\
hrswq16&0.007&0.631&7.157&0.000&0.007&0.666&6.998&0.000\\
work208&0.001&0.991&11.675&0.000&0.001&0.985&11.081&0.000\\
			\hline\hline
		\end{tabular}
	\end{center}
	\par
	Note: Outcomes `earny4', `earnq16', `hrswq16', and `work208' are weekly earnings in the fourth year, earnings and hours worked per week in quarter 16, and a binary employment indicator four years after assignment (i.e.\ in week 208). `stat' denotes the test statistic.
\end{table}

It is worth mentioning that the concavities in the outcome-treatment associations spotted in the previously mentioned empirical applications suggest decreasing marginal effects when increasing the treatment. In our testing context, this implies that weakly negative monotonicity should be more clearly rejected for lower rather than higher ranges of treatment values by our method. To verify this suspicion, we in a second step partition the treatment support into three sets of $[40,1000]$, $[1000,2000]$, and $[2000,3000]$ and run the tests separately within each set.

\begin{table}[htp]
	\begin{center}
		\caption{Test statistic and p-value, $1000\leq T \leq 2000$}\label{Table:1000_2000}
		\begin{tabular}{ccccccccc}
			\hline\hline
			&\multicolumn{4}{c}{N=40, $t_1>t_2$}&\multicolumn{4}{c}{N=50, $t_1>t_2$}\\
			\cmidrule(lr){2-5} \cmidrule(lr){6-9}
			$H_0:$ &\multicolumn{2}{c}{$\mu(t_1)\geq\mu(t_2)$}& \multicolumn{2}{c}{$\mu(t_1)\leq\mu(t_2)$}
			&\multicolumn{2}{c}{$\mu(t_1)\geq\mu(t_2)$}& \multicolumn{2}{c}{$\mu(t_1)\leq\mu(t_2)$} \\
			$Y$& stat & p-value & stat & p-value& stat & p-value & stat & p-value\\
			\hline
            earny4&0.075&0.672&0.485&0.206&0.076&0.631&0.468&0.238\\
earnq16&0.554&0.200&0.029&0.860&0.524&0.207&0.038&0.814\\
hrswq16&0.563&0.183&0.088&0.580&0.552&0.194&0.088&0.552\\
work208&0.419&0.226&0.232&0.393&0.415&0.225&0.264&0.346\\
			\hline\hline
		\end{tabular}
	\end{center}
		\par
		Note: Outcomes `earny4', `earnq16', `hrswq16', and `work208' are weekly earnings in the fourth year, earnings and hours worked per week in quarter 16, and a binary employment indicator four years after assignment (i.e.\ in week 208). `stat' denotes the test statistic.
\end{table}

Table \ref{Table:40_1000} presents the results for  $T \in [40,1000]$. None of the tests rejects weakly positive monotonicity at any conventional level of significance, while all tests strongly reject weakly negative monotonicity. For the intermediate treatment range of $[1000,2000]$ considered in Table \ref{Table:1000_2000}, however, neither positive nor negative monotonicity is ever rejected at the 10\% level of statistical significance. This implies that marginal treatment effects are generally less positive than for lower values of $T$. The same findings apply to the highest treatment bracket $[2000,3000]$, where all tests yield p-values which are beyond conventional levels of significance. Summing up, our empirical findings are consistent with a concave mean potential outcome-treatment dependence, implying that initially strongly positive marginal treatment effects decrease as the treatment value  considered (hours in training) increases. A potential explanation for the concavity could be that individuals attending more training in the first year might be induced to attain more education also in the following years rather than to participate in the labor market.

	\begin{table}[htp]
	\begin{center}
		\caption{Test statistic and p-value, $2000\leq T \leq 3000$}\label{Table:2000_3000}
		\begin{tabular}{ccccccccc}
			\hline\hline
			&\multicolumn{4}{c}{N=40, $t_1>t_2$}&\multicolumn{4}{c}{N=50, $t_1>t_2$}\\
			\cmidrule(lr){2-5} \cmidrule(lr){6-9}
			$H_0:$ &\multicolumn{2}{c}{$\mu(t_1)\geq\mu(t_2)$}& \multicolumn{2}{c}{$\mu(t_1)\leq\mu(t_2)$}
			&\multicolumn{2}{c}{$\mu(t_1)\geq\mu(t_2)$}& \multicolumn{2}{c}{$\mu(t_1)\leq\mu(t_2)$} \\
			$Y$& stat & p-value & stat & p-value& stat & p-value & stat & p-value\\
			\hline
             earny4&0.029&0.600&0.487&0.211&0.023&0.641&0.472&0.199\\
earnq16&0.008&0.889&0.591&0.178&0.007&0.876&0.543&0.210\\
hrswq16&0.132&0.353&0.205&0.353&0.149&0.346&0.176&0.385\\
work208&0.465&0.229&0.020&0.723&0.457&0.231&0.014&0.758\\
			\hline\hline
		\end{tabular}
	\end{center}
	\par
	Note: Outcomes `earny4', `earnq16', `hrswq16', and `work208' are weekly earnings in the fourth year, earnings and hours worked per week in quarter 16, and a binary employment indicator four years after assignment (i.e.\ in week 208). `stat' denotes the test statistic.
\end{table}

\section{Testing Monotonicity Conditional on Covariates}\label{cond}
In this section, we adapt our method to testing monotonicity with conditional (rather than unconditional) mean potential outcomes given observed covariates $X$. In this case, the null hypothesis considered corresponds to
\begin{align}
H_{0}:~ \mu(t_1,x)\geq \mu(t_2,x),~~\text{for all}~t_{1}\geq t_{2},~\text{for $t_1,t_2\in[0,1]$ and }~{x\in\mathcal{X}},\label{eq: null-x}
\end{align}
where $\mu(t,x)=E[Y(t)|X=x]$ is the conditional average of the potential outcome function or the average dose-response function.
For simplicity and without loss of generality, we henceforth assume that $X$ is a scalar with $\mathcal{X}=[0,1]$.
By Lemma 2.1 of \cite{HsuShen2020}, $H_0$ in (\ref{eq: null-x}) is equivalent to
\begin{align}
&\int_{{x}}^{{x}+q^{-1}}\int_{t_2}^{t_2+q^{-1}} \mu(s,\tilde{x}) \cdot h(s,\tilde{x}) dsd\tilde{x}
\cdot \int_{{x}}^{{x}+q^{-1}}\int_{t_1}^{t_1+q^{-1}}  h(s,\tilde{x}) dsd\tilde{x} -\notag\\
&~~~~~~
\int_{{x}}^{{x}+q^{-1}}\int_{t_1}^{t_1+q^{-1}} \mu(s,\tilde{x})\cdot  h(s,\tilde{x}) dsd\tilde{x}  \cdot
\int_{{x}}^{{x}+q^{-1}}\int_{t_2}^{t_2+q^{-1}}  h(s,\tilde{x}) dsd\tilde{x} \leq 0 \label{eq: inequality-x}
\end{align}
for any $q=2,\cdots,$ and for any $t_1\geq t_2$ such
that $q\cdot t_1,q\cdot t_2, q\cdot x  \in \{0,1,2,\cdots,q-1\}$.
Define $h(t,x)=f(x)$ to be the density function of $X$. Following Lemma \ref{Lid} and (\ref{eq: DML nu iden}), we have for $r>0$,
\begin{align*}
&\int_{{x}}^{{x}+r}\int_t^{t+r} \mu(s,\tilde{x}) h(s,\tilde{x}) ds d\tilde{x} =
E\left[ \frac{Y}{p(T,X)} \cdot  1(T\in[t,t+r])\cdot 1(X\in[x,x+r])\right]
\\
&~~~~~=E\left[\left\{E\Big[ \frac{Y{\bf 1}(T\in [t, t+r])}{p(T,X)}\bigg|X\Big] +  \frac{Y - \gamma(T,X)}{p(T,X)}{{\bf 1}(T\in[t, t+r])}\right\}1(X\in[x,x+r]) \right],\\
&
\int_{{x}}^{{x}+r}\int_t^{t+r} h(s,\tilde{x}) ds d\tilde{x} =
E\left[ 1(X\in[x,x+r])\right].
\end{align*}
For $\ell_x=(t_1,t_2,x,q^{-1})\in[0,1]^3\times (0,1]$, we let
\begin{align}
{\cal L}_x=
\Big\{\ell_x=(t_1,t_2,x,q^{-1}):~q\cdot(t_1,t_2,x)&\in\{0,1,2,\cdots,q-1\}^{3},~t_1> t_2\notag \\
&~~~~\text{, and } q=2,3,\cdots\Big\}.\label{eq: Gc-cube-x}
\end{align}
Similar to (\ref{eq: DML nu iden}), for each $\ell_x$, we define
\begin{align*}
&\nu_1(\ell_x)=E\left[\left\{E\Big[ \frac{Y{\bf 1}(T\in [t_1, t_1+q^{-1}])}{p(T,X)}\bigg|X\Big] +  \frac{Y - \gamma(T,X)}{p(T,X)}{{\bf 1}(T\in[t_1, t_1+q^{-1}])}\right\}1(X\in[x,x+q^{-1}]) \right],\\
&\nu_2(\ell_x)=E\left[\left\{E\Big[ \frac{Y{\bf 1}(T\in [t_2, t_1+q^{-1}])}{p(T,X)}\bigg|X\Big] +  \frac{Y - \gamma(T,X)}{p(T,X)}{{\bf 1}(T\in[t_1, t_1+q^{-1}])}\right\}1(X\in[x,x+q^{-1}]) \right].
\end{align*}
This permits establishing the following lemma.

\begin{lemma}\label{lemma: H0 transformed-x}
	Suppose Assumption \ref{assu: unconfoundedness} holds.  Assume that $\mu(t,x)$ is continuous in $t$ for all $x\in[0,1]$.  Then $H_0$ in (\ref{eq: null-x}) is equivalent to
	\begin{align}
	H_0':&~\nu(\ell_x)=\nu_2(\ell_x)-\nu_1(\ell_x)\leq 0~\text{for any  $\ell_x=(t_1,t_2,x,q^{-1})\in\mathcal{L}_x$}.\label{eq: null-2-x}
	\end{align}
\end{lemma}

Similar to Section \ref{testDML}, we estimate $\nu(\ell_x)$ with $\ell_x=(t_1,t_2,x,q^{-1})$ as the following:

\begin{changemargin}{0.6cm}{0cm}
	\begin{itemize}
		\item[Step 1.]
		(Cross-fitting) For some fixed $K \in \{2,...,n\}$, a $K$-fold cross-fitting partitions the observation indices into $K$ distinct groups $I_k$, $k=1,..., K$, such that the sample size of each group is the largest integer smaller than $n/K$.
		For $k \in \{1,...,K\}$, the estimators $\hat \gamma_k(t, x)$ and $\hat p_k(t , x)$ use observations not in $I_k$ and satisfy Assumption~\ref{A1st} below.
		
		\item[Step 2.]
		(Double robustness) The DML estimator is defined as
		\begin{align*}
		&\hat\nu_{DML}(\ell_x) = \hat\nu_{2,DML}(\ell_x)-\hat\nu_{1,DML}(\ell),~\text{where for $j=1$ and 2,}\\
		&\hat\nu_{j,DML}(\ell_x)\\
		&=
		\frac{1}{K}\sum_{k=1}^K\frac{1}{n_k}\sum_{i\in I_k} \Big\{\int_{t_j}^{{t_j}+q^{-1}} \hat \gamma_k(s,X_i) ds \\
		&~~~~~~~~~~~~~~~~~~~~~~~~~+ \frac{Y_i - \hat \gamma_k(T_i,X_i)}{\hat p_k(T_i,X_i)}{{\bf 1}(T_i\in[t_j, t_j+q^{-1}])}\Big\}1(X_i\in[x,x+q^{-1}]),
		\end{align*}
		and $\int_{t_j}^{t_j+q^{-1}} \hat \gamma_k(s,X_i) ds$ is approximated as in Section \ref{testDML}.
	\end{itemize}
\end{changemargin}

\medskip

Similar to Lemma \ref{TIF}, we can show that uniformly over $\ell_x\in\mathcal{L}_x$,
\begin{align}
\sqrt{n}(\hat{\nu}_{DML}(\ell_x)-\nu(\ell_x))
&=\frac{1}{\sqrt{n}}\sum_{i=1}^n\phi_{\ell_x,DML}(Y_i,T_i,X_i)+o_p(1),\label{eq: influ nu-x}
\end{align}
where
\begin{align*}
&\phi_{\ell_x,DML}(Y_i,T_i,X_i)=\phi_{2,\ell_x,DML}(Y_i,T_i,X_i)-\phi_{1,\ell_x,DML}(Y_i,T_i,X_i),~\text{and for $j=1$ and 2, }\\
&\phi_{j,\ell_x,DML}(Y,T,X)\\
&=1(X\in[x,x+q^{-1}])\Big(E\left[ \frac{Y{\bf 1}(T\in [t_j, t_j+q^{-1}])}{p(T,X)}\bigg|X\right] +  \frac{Y - \gamma(T,X)}{p(T,X)}{{\bf 1}(T\in[t_j, t_j+q^{-1}])}  \Big)- \nu_j(\ell_x).
\end{align*}
Let $\hat{\phi}_{\ell_x,DML}(Y,T,X)$ be the estimated influence function similar to (\ref{eq: estimated influ DML}) and
let $\hat{\sigma}^2_{\nu,DML}(\ell_x)=K^{-1}\sum_{k=1}^K{n^{-1}_k}\sum_{i\in I_k} \hat{\phi}^2_{\ell_x,DML}(Y_i,T_i,X_i)$ which will be a consistent estimator for the asymptotic variance of $\sqrt{n}(\hat{\nu}_{DML}(\ell_x)-{\nu}(\ell_x))$ under proper regularity conditions.  Furthermore, let $\hat{\sigma}_{\nu,\epsilon,DML}(\ell_x)=\max\{\hat{\sigma}_{\nu,DML}(\ell_x), \epsilon\cdot \hat{\sigma}_{\nu,DML}(0,1/2, 0,1/2)\}$. The  Cram\'{e}r-von Mises test
statistic is defined as
\begin{align}
\widehat{T}_{x,DML}= \sum_{\ell_x\in\mathcal{L}_x} \max\Big\{ \sqrt{n}\frac{\hat{\nu}_{DML}(\ell_x)}{\hat{\sigma}_{\nu,\epsilon,DML}(\tau,\ell_x)},
0 \Big\}^2 Q(\ell_x),
\label{eq: test statistic-x}
\end{align}
where $Q$ is a weighting function such that $Q(\ell_x)>0$ for all  $\ell_x \in \mathcal{L}_x$ and
$\sum_{\ell_x\in\mathcal{L}_x}Q(\ell_x)<\infty$.
The simulated process is constructed as
\begin{align} \label{eq: simulated process-mu2}
&\widehat{\Phi}_{\nu,x,DML}^u(\ell_x)=\frac{1}{\sqrt{n}}\sum_{i=1}^nU_i\cdot\hat{\phi}_{\ell_x,DML}(Y_i,T_i,X_i).
\end{align}
The GMS simulated critical value is given by
\begin{align*}
\hat{c}^{\eta}_{x,DML}(\alpha)& =\sup\left\{q\Big|P^u
\left(\sum_{\ell_x\in \mathcal{ L}_x} \max\Big\{\frac{\widehat{\Phi}_{\nu,x,DML}^u(\ell_x)}{\hat \sigma_{\nu,\epsilon,DML}(\ell_x)}+\hat{\psi}_{\nu,DML}(\ell_x),0\Big\}Q(\ell_x) \leq q\right)\leq 1-\alpha +\eta \right\}+\eta,\\
\hat{\psi}_{\nu,DML}(\ell_x)&=-B_n \cdot 1\left(\sqrt{n}\cdot \frac{\hat{\nu}_{DML}(\ell_x)}{\hat \sigma_{\nu,\epsilon,DML}(\ell_x)} <-a_n\right).
\end{align*}
Finally, the decision rule is given by
\begin{align*}
\text{Reject $H'_0$ if $\widehat{T}_{x,DML}>\hat{c}^{\eta}_{x,DML}(\alpha)$. }
\end{align*}
The size and power properties are similar to the unconditional potential outcome cases and the details are omitted for brevity.

\section{Conclusion}\label{conclusion}
In this paper, we propose Cram\'{e}r-von Mises-type tests for testing whether a mean potential outcome is weakly monotonic in a continuously distributed treatment under a weak unconfoundedness assumption.
To flexibly employ nonparametric or machine learning estimators in the presence of possibly high-dimensional nuisance parameters, we propose a double debiased machine learning estimator for the moments entering the test.
Furthermore, we extend our method to testing monotonicity conditional on observed covariates.
We also investigate the test's finite sample behavior in a simulation study and find it to perform decently under our suggested choices of tuning parameters.

As an empirical illustration, we apply our test to the Job Corps study, investigating the associations of several labor market outcomes (earnings, employment, and hours worked) with hours in training as treatment. We find that an increase in the treatment does either increase or at least not reduce the outcome.
When splitting the treatment range into subsets, our testing results are consistent with a concave mean potential outcome-treatment dependence, implying that initially stronger marginal treatment effects decrease as the treatment value (i.e.\ hours already spent in training) increases.

\newpage

\begin{center}
{\Large APPENDIX}
\par\end{center}

\appendix

\section{Proof of Lemma \ref{Lid}}
Under  Assumption \ref{assu: unconfoundedness}, \cite{HiranoImbens2004} show that
\begin{align*}
\mu(t) = E\left[ E\left[ Y|T=t, p(t,X) \right]  \right] = \int_\mathcal{X} E\left[ Y|T=t, p(T,X) = p(t,X) \right]  f(x)dx.
\end{align*}
Then
\begin{align*}
\int_t^{t+r} \mu(t) h(s) ds &= \int_t^{t+r}   \int_\mathcal{X} E\left[ Y|T=s, p(T,X) = p(s,X) \right]  f(x) h(s) dxds  \\
&= E\left[     E\left[ Y|T, p(T,X)\right]  \frac{f(X)h(T) 1(T\in[t, t+r])}{f_{TX}(T,X)} \right] \\
&= E\left[     E\left[ Y|T, p(T,X) \right]  \frac{h(T)1(T\in[t, t+r])}{p(T,X)}  \right] \\
&= E\left[    \frac{Y}{p(T,X)} h(T)1(T\in[t, t+r]) \right].
\end{align*}
\hfill $\square$

\section{Appendix for Section \ref{testDML}}\label{app: dml test}
\noindent {\bf Proof of Lemma~\ref{TIF}}:\\
We give an outline of deriving the asymptotically linear representation, following \cite{CEINR}.
Let $\nu(t,r)=\int_t^{t+r} \mu(s) ds$ and
\begin{align*}
\hat\nu_{DML}(t,r)&=
\frac{1}{n}\sum_{i=1}^n \left\{\int_{t}^{{t}+r} \hat \gamma_k(s,X_i) ds + \frac{Y_i - \hat \gamma_k(T_i,X_i)}{\hat p_k(T_i,X_i)}{{\bf 1}(T_i\in[t, t+r])}\right\},
\end{align*}
To show Lemma \ref{TIF}, it is sufficient to show that uniformly over $(t,r)\in[0,1]^2$,
\begin{align}
&\sqrt{n}(\hat\nu_{DML}(t,r)-\nu(t,r))\notag\\
=&
\frac{1}{\sqrt{n}}\sum_{i=1}^n
E\left[ \frac{Y{\bf 1}(T\in [t, t+r])}{p(T,X)}\bigg|X\right] +  \frac{Y - \gamma(T,X)}{p(T,X)}{{\bf 1}(T\in[t, t+r])}-\nu(t,r) +o_p(1).
\label{app eq: phi}
\end{align}

For notational ease, let $Z_i=(Y_i,X_i,T_i)$, $\gamma_i \equiv \gamma(T_i, X_i)$ and $\lambda_i \equiv \lambda(T_i, X_i) = 1/f_{T|X}(T_i|X_i)$.
Let the doubly robust moment function in equation (\ref{EDR}) be
\begin{align*}
&\phi_{(t,r)}(Z_i, \gamma, \lambda)\\
\equiv &E\left[ Y{\bf 1}(T\in [t, t+r])\lambda(T,X)\big|X= X_i\right] - \nu(t,r) +  (Y_i - \gamma(T_i,X_i))\lambda(T_i,X_i){{\bf 1}(T_i\in[t, t+r])}.
\end{align*}
Let $Z_k^c$ denote the observations $Z_i$ for $i \neq I_k$
and $\hat \gamma_{ik} = \hat r_k(T_i, X_i)$ using $Z_k^c$ for $i \in I_k$.
We decompose the remainder term
\begin{align}
&\sqrt{n}\frac{1}{n}\sum_{i=1}^n \left\{ \hat \phi_{(t,r)}(Z_i, \hat \gamma, \hat \lambda) - \phi_{(t,r)}(Z_i, \gamma, \lambda) \right\} \notag \\
=&\ \frac{1}{\sqrt{n}} \sum_{k=1}^K\sum_{i\in I_k} \bigg\{ \int_t^{t+r}\left( \hat \gamma_k(s, X_i) - \gamma(s, X_i) \right) d{s}- E\left[ \int_t^{t+r}\left( \hat \gamma_k(s, X_i) - \gamma(s, X_i) \right) d{s} \bigg|Z_k^c\right]  \tag{R1-1}\label{ER11}\\
& + {\bf 1}(T_i \in [t, t+r])\lambda_i (\gamma_i - \hat \gamma_{ik}) - E\big[{\bf 1}(T_i \in [t, t+r])\lambda_i (\gamma_i -\hat \gamma_{ik})\big|Z_k^c\big]  \tag{R1-2} \label{ER12} \\
& + {\bf 1}(T_i \in [t, t+r])(\hat \lambda_{ik} - \lambda_i)(Y_i - \gamma_i) - E\big[{\bf 1}(T_i \in [t, t+r])(\hat \lambda_{ik} - \lambda_i)(Y_i - \gamma_i) \big|Z_k^c\big]\bigg\}
\tag{R1-3} \label{ER13} \\
& + \sqrt{n} \bigg\{ E\left[ \int_t^{t+r}(\hat \gamma_k(s, X_i) - \gamma(s, X_i))ds\bigg|Z_k^c \right]
- E\left[ {\bf 1}(T_i \in [t, t+r]) \lambda_i (\hat\gamma_{ik} - \gamma_i)|Z_k^c\right] \notag\\
& + E[ (\hat \lambda_{ik} - \lambda_i) {\bf 1}(T_i \in [t, t+r]) (Y_i - \gamma_i)|Z_k^c] \bigg\}
\tag{R1-DR} \label{ER1DR} \\
& - \frac{1}{\sqrt{n}}\sum_{k=1}^K\sum_{i\in I_k}  {\bf 1}(T_i \in [t, t+r])\big(\hat \lambda_{ik} - \lambda_i\big)\big(\hat \gamma_{ik} - \gamma_i\big). \tag{R2} \label{ER2} 
\end{align}
The remainder terms (\ref{ER11}), (\ref{ER12}) and (\ref{ER13}) are stochastic equicontinuous terms that are controlled to be $o_p(1)$ by
the mean-squared consistency conditions in Assumption~\ref{A1st}(i) and cross-fitting.
The second-order remainder term (\ref{ER2}) is controlled by Assumption~\ref{A1st}(ii).

Note that we can express
\begin{align}
\int_t^{t+r} \gamma_k(s, X_i) ds = E\left[\frac{\gamma(T,X){\bf 1}(T\in[t, t+r])}{p(T,X)} \bigg| X=X_i\right].
\label{ER0DR}
\end{align}
By the law of iterated expectations, $E\Big[ \int_t^{t+r} \big( \hat \gamma_k(s, X) - \gamma(s, X)\big) ds \Big|Z_k^c\Big] =  E\Big[  \lambda(T,X) \big( \hat \gamma_k(T, X) -  \gamma(T,X) \big) {\bf 1}(T\in [t, t+r])\Big|Z_k^c\Big]$.
So (\ref{ER1DR}) is zero.

\medskip

The approximation error of the Riemann sum is
\begin{align*}\big| M^{-1}\sum_{m=1}^M  \hat \gamma_k(t_m,X_i) - \int_t^{t+r} \hat \gamma_k(s,X_i) ds \big| \leq
M^{-1}\sum_{m=1}^M \big| \hat \gamma_k(t_m,X_i) - \hat \gamma_k(t_{m-1},X_i) \big|
= O_p(M^{-1}),
\end{align*} by Assumption~\ref{A1st}(iii).
By the condition $\sqrt{n}/M \rightarrow 0$, the approximation error is asymptotically ignorable.

\bigskip

To show (\ref{ER11}), (\ref{ER12}) and (\ref{ER13}) are $o_p(1)$ uniformly over $\ell$,
we show these terms weakly converge to Gaussian processes indexed by $\ell$ with zero covariance kernel.
It suffices to show the results with ${\bf 1}(T_i \leq t)$ replacing ${\bf 1}(T_i \in [t, t+r])$.
We apply the functional central limit theorem in Theorem 10.6 in \cite{Pollard}.
Following the notation in \cite{Pollard},
for any $\omega$ in the probability space $\Omega$ and for $i \in I_k$,
define $f_{i}(t) = f_{i}(\omega, t) = {\bf 1}(T_i \leq t) \lambda_i (\hat\gamma_{ik} - \gamma_{i})$ for (\ref{ER12}) and $f_{ni}(t) = f_{i}(t)/\sqrt{n}$.
Due to cross-fitting, the processes from the triangular array $\{f_{ni}(t)\}$ given $Z_k^c$ are independent within rows.
Let $n_k = \sum_{i=1}^n {\bf 1}(i\in I_k)$.
Since $K$ is fixed, $n/n_k = O(1)$.
We verify the conditions in Theorem 10.6 in \cite{Pollard}.
\begin{itemize}
	\item[(i)]
	$\{{\bf 1}(T_i \leq t): t\in [0,1], i \in I_k\}$ is manageable since it is monotone increasing in $t$ (p.221 in \cite{Kosorok}).
	The triangular array processes $\{f_{ni}(t)\}$ are manageable with respect to the envelopes $F_{ni} = \big| \lambda_i (\hat \gamma_{ik}- \gamma_i) \big| \big/\sqrt{n}$.
	$F_{n_k} = (F_{n1},..., F_{nn_k})'$ is a $R^{n_k}$-valued function on the underlying probability space.
	
	\item[(ii)]
	Let $\mathsf{X}_n(t)  = \mathsf{X}_n(\omega, t) = \sum_{i\in I_k} \left(f_{ni}(t) - E\left[f_{ni}(t)\big|Z_k^c\right]\right)$.
	By construction and independence of $Z_k^c$ and $z_i, i\in I_k$, $E[f_{ni}(t)|Z_k^c] = 0$ and $E[f_{ni}(t)f_{nj}(t)|Z_k^c] = 0$ for $i,j \in I_k$.
	For $i \in I_k$, $E[f_{i}(t)^2|Z_k^c] = O_p(\|\hat \gamma_{ik} - \gamma_i)\|_2^2) = o_p(1)$ by Assumption~\ref{A1st}(i) and (iv).
	Let $s\leq t \in[0,1]$, without loss of generality.
	$H(s,t) = \lim_{n\rightarrow\infty} E\left[\mathsf{X}_n(s) \mathsf{X}_n(t) \big|Z_k^c \right]
	= \lim_{n\rightarrow\infty} E\left[{\bf 1}(T\in(s,t]) \lambda_i^2 (\hat\gamma_{ik}-\gamma_i)^2\big|Z_k^c\right]
	= 0$.
	
	\item[(iii)] By the argument in (ii), $H(t,t) = 0$.
	
	\item[(iv)] For each $\epsilon > 0$,
	\begin{align*}
	\sum_{i\in I_k}E[F_{ni}^2\{F_{ni} \geq \epsilon\}|Z_k^c] \leq
	\sum_{i\in I_k}E[F_{ni}^2|Z_k^c] = O_p\left(\|\hat\gamma-\gamma\|_2^2]\right) = o_p(1).
	\end{align*}
	
	\item[(v)]
	For any $s<t$,
	\begin{align*}
	\rho_n(s,t) &= \left(\sum_{i \in I_k} E\left[\left| f_{ni}(s) - f_{ni}(t) \right|^2\Big|Z_k^c\right]\right)^{1/2}=
	\left(E\left[\left| {\bf 1}(T_i \in (s,t]) \lambda_i (\hat\gamma_{ik}-\gamma_i)\right|^2\Big|Z_k^c\right]\right)^{1/2}\\
	&= O_p\left(\|\hat \gamma_k-\gamma\|_2\right)= o_p(1)
	\end{align*}
	and the last equality holds by Assumption~\ref{A1st}(i).
	Hence, $\rho(s,t)  = \lim_{n\rightarrow\infty}\rho_n(s,t) = 0$.
	The condition (v) holds: for all deterministic sequences $\{s_n\}$ and $\{t_n\}$, if $\rho(s_n, t_n) \rightarrow 0$ then $\rho_n(s_n, t_n)\rightarrow 0$.
\end{itemize}

Then Theorem 10.6 in \cite{Pollard} implies that the finite dimensional distributions of $\mathsf{X}_n$ have Gaussian limits, with zero means and covariances given by $H$.
Therefore, $\mathsf{X}_n = o_p(1)$ uniformly over $t\in[0,1]$.

\bigskip

The analogous results also hold for
$f_{i}(t) =
{\bf 1}(T_i \leq t)(\hat \lambda_{ik} - \lambda_i)(Y_i - \gamma_i)$ in (\ref{ER13}).
In particular, for (\ref{ER13}),
$E[f_{ni}(t)^2|Z_k^c] =
O_p\left( \|\hat\lambda_k - \lambda\|_2^2\right) =o_p(1)$ by the smoothness condition and Assumption~\ref{A1st}(i).

\medskip

For (\ref{ER11}), define
$f_{i}(t) =
\int_0^t\left( \hat \gamma_k(s, X_i) - \gamma(s, X_i) \right) d{s}$.
By (\ref{ER0DR}),
\begin{align*}
E[f_{i}(t)^2|Z_k^c] &\leq
\int \left(
E\left[\frac{(\hat\gamma_k(T,X) - \gamma(T,X))}{p(T,X)} {\bf 1}(T \leq t) \Big| X=X_i\right]
\right)^2 f_X(X_i) dX_i \\
&
\leq
\int
E\left[\left(\frac{\hat\gamma_k(T,X) - \gamma(T,X)}{p(T,X)} {\bf 1}(T\leq t)\right)^2 \Big| X=X_i\right]
f_X(X_i) dX_i\\
&
= \int\int\left(\frac{\hat\gamma_k(T_i,X_i) - \gamma(T_i,X_i)}{p(T_i,X_i)} \right)^2 {\bf 1}(T_i\leq t) f_{T|X}(T_i|X_i)dT_i dX_i\\
&
= O_p\left(
\int\int\left( \hat\gamma_k(T_i,X_i) - \gamma(T_i,X_i) \right)^2 f_{T|X}(T_i|X_i)dT_i dX_i
\right)\\
& = o_p(1)
\end{align*}
and the last equality holds by Assumption~\ref{A1st}(i).

\medskip

For (\ref{ER2}),
\begin{align}
&E\bigg[ \sup_\ell \Big| n^{-1/2} \sum_{i \in I_k} {\bf 1}(T_i \in [t, t+r])(\hat \lambda_{ik} - \lambda_i)(\gamma_i - \hat \gamma_{ik})  \Big|  \bigg|Z_k^c\bigg] \notag\\
&\leq
\sqrt{n} \int_\mathcal{X}\int_\mathcal{T} \sup_\ell {\bf 1}(T_i \in [t, t+r])\Big| (\hat \lambda_{ik} - \lambda_i)(\gamma_i - \hat \gamma_{ik})   \Big| f_{TX}(T_i,X_i)dT_i dX_i \notag\\
&\leq \sqrt{n}
\Big(\int_\mathcal{X} \int_\mathcal{T} (\hat\lambda_{ik} - \lambda_i)^2f_{TX}(T_i, X_i) dT_idX_i \Big)^{1/2}
\Big(\int_\mathcal{X} \int_\mathcal{T} (\hat\gamma_{ik} - \gamma_i)^2f_{TX}(T_i, X_i) dT_idX_i \Big)^{1/2}  \notag\\
&\stackrel{p}{\longrightarrow} 0
\label{ER2pf}
\end{align}
by Cauchy-Schwartz inequality and Assumption~\ref{A1st}(ii).
By the conditional Markov and triangle inequalities, (\ref{ER2})$\stackrel{p}{\longrightarrow} 0$ uniformly over $\ell$.

\medskip

By the triangle inequality, we obtain the asymptotically linear representation
\begin{align*}
n^{-1/2}\sum_{i=1}^n \big( \hat \phi_{t,r}(Z_i, \hat \gamma, \hat \lambda) - \phi_{t,r}(Z_i, \gamma, \lambda) \big) = o_p(1),
\end{align*}
and (\ref{app eq: phi}) follows.

\medskip
Then by the fact that $\nu(\ell)=\nu(t_1,q^{-1})-\nu(t_2,q^{-1})$, then it follows that uniformly over $\ell\in\mathcal{L}$,
\begin{align*}
\sqrt{n}(\hat\nu_{DML}(\ell) - \nu(\ell)) = n^{-1/2}\sum_{i=1}^n\phi_{\ell,DML}(Y_i,T_i,X_i)+o_p(1),
\end{align*}
and this shows the first half of Lemma \ref{TIF}.

For the second part, similar to \cite{HsuLiuShi2019}, it is straightforward to see that
$\{\phi_{\ell,DML}(Y,T,X): \ell\in\mathcal{L}\}$ is a VC class of functions and by functional central limit theorem of \cite{Pollard}, it follows that  $\sqrt{n}(\hat{\nu}_{DML}(\cdot)-\nu(\cdot))\Rightarrow \Phi_{h_{DML}}(\cdot)$ where
$\Phi_{h_{DML}}(\cdot)$ is a Gaussian process
with variance-covariance kernel $h_{DML}(\ell_1,\ell_2)=E[\phi_{\ell_1,DML}(Y,T,X) \phi_{\ell_2,DML}(Y,T,X)]$.  This completes the proof of Lemma \ref{TIF}.
\hfill $\square$

\bigskip

\begin{lemma}\label{lemma: simulated-np}
Suppose the Assumptions Assumptions \ref{assu: unconfoundedness}, \ref{A1st} and  \ref{assu: U-1} hold.  Then, $\sup_{\ell\in\mathcal{L}}|\hat{\sigma}_{\nu,DML}(\ell)-{\sigma}_{\nu,DML}(\ell)|\stackrel{p}{\rightarrow} 0$ where ${\sigma}^2_{\nu,DML}(\ell)=E[\phi^2_{\ell,DML}]$, and $\widehat{\Phi}^u_{\nu,DML}{\Rightarrow} \Phi_{h_{DML}}$ conditional on sample path with probability approaching one.
\end{lemma}

\noindent {\bf Proof of Lemma \ref{lemma: simulated-np}:}\\
The fact that $\{\phi_{\ell,DML}: \ell\in\mathcal{L}\}$ is a VC type class of functions implies that
$\{\phi^2_{\ell,np}: \ell\in\mathcal{L}\}$ is also a VC type.  In addition, given that $E[\bar{\phi}^{2+\delta}_{np}]<\infty$, we have by the uniform weak law of large numbers that
$\sup_{\ell\in\mathcal{L}}|\tilde{\sigma}^2_{\nu,DML}(\ell)-{\sigma}^2_{\nu,DML}(\ell)|\stackrel{p}{\rightarrow} 0$,
where $\tilde{\sigma}^2_{\nu,DML}(\ell)=n^{-1}\sum_{i=1}^n \phi^2_{\ell,DML}(Y_i,T_i,X_i)$. By Assumption \ref{A1st}, we have that $\sup_{\ell\in\mathcal{L}}|\tilde{\sigma}^2_{\nu,DML}(\ell)-\hat{\sigma}^2_{\nu,DML}(\ell)|\stackrel{p}{\rightarrow} 0$.  Then the first part follows.  The proof of the second part follows from the standard arguments for the multiplier bootstrap such as Lemma 4.1 of \cite{Hsu2017} and is omitted for the sake of brevity.
\hfill $\square$

\bigskip

\noindent {\bf Proof of Theorem \ref{thm: size and power}:}\\
The proof of Theorem \ref{thm: size and power} follows from the same arguments as Theorem 5.1 of \cite{Hsu2017} once Lemmas \ref{TIF} and \ref{lemma: simulated-np} are established and is omitted for the sake of brevity.
\hfill $\square$

\newpage
\bibliographystyle{chicago} 
\bibliography{CTE_mono_DML}         

\end{document}